\documentclass[twocolumn,showpacs,preprintnumbers,amsmath,amssymb,superscriptaddress,10pt,aps,prb]{revtex4-2}
\usepackage{epsfig,amsopn}
\usepackage{graphicx}
\usepackage{epstopdf}
\usepackage{tikz}
\usetikzlibrary{shapes, arrows.meta, positioning}
\usepackage[mathscr]{euscript}
\DeclareSymbolFont{rsfs}{U}{rsfs}{m}{n}
\DeclareSymbolFontAlphabet{\mathscrsfs}{rsfs}

\usepackage{amsmath,amssymb}
\usepackage{amsthm}
\usepackage{enumerate}
\usepackage{xcolor}
\usepackage[normalem]{ulem}
\usepackage[colorlinks=true,allcolors=blue]{hyperref}

\usepackage{ulem}
\newcommand\bs{\boldsymbol}

\newcommand\mc{\mathcal}

\newcommand{\moire}{moir\'e }
\newcommand{\Moire}{Moir\'e }

\usepackage[defaultcolor=blue]{changes}
\definechangesauthor[name=Arijit, color=orange]{ak}
\definechangesauthor[name=Herb, color=blue]{haf}

\begin{document}
\title{Skyrmion Stripes in Twisted Double Bilayer Graphene}
\author{Debasmita Giri}
\affiliation{Department of Physics, Indian Institute of Technology Kanpur, Kanpur 208016, India}
\author{Dibya Kanti Mukherjee}
\affiliation{Laboratoire  de  Physique  des  Solides,  CNRS  UMR  8502, Universit\`{e} Paris-Saclay,  91405  Orsay  Cedex,  France}
\affiliation{Department of Physics, Indiana University, Bloomington, IN 47405}
\author{H.A. Fertig}
\affiliation{Department of Physics, Indiana University, Bloomington, IN 47405}
\affiliation{Quantum Science and Engineering Center, Indiana University, Bloomington, IN, 47408}
\affiliation{Instituto de Ciencia de Materiales de Madrid, (CSIC), Cantoblanco, 28049, Madrid, Spain}
\author{Arijit Kundu}
\affiliation{Department of Physics, Indian Institute of Technology Kanpur, Kanpur 208016, India}

\begin{abstract}
Two dimensional \moire systems have recently emerged as a platform in which the interplay between topology and strong correlations of electrons play out in non-trivial ways. Among these systems, twisted double bilayer graphene (TDBG) is of particular interest as its topological properties may be tuned via both twist angle and applied perpendicular electric field.  In this system, energy gaps are observed at half filling of particular bands, which can be associated with correlated spin polarized states. In this work, we investigate the fate of these states as the system is doped away from this filling. We demonstrate that, for a broad range of fractional fillings, the resulting ground state is partially valley polarized, and supports multiple broken symmetries, including a textured spin order indicative of skyrmions, with a novel {\it stripe} ordering that spontaneously breaks $C_3$ symmetry.  Experimental signatures of this state are discussed.
\end{abstract}

\maketitle

\section{Introduction.}
Moir{\'e} materials have attracted intense interest since
the experimental discovery of correlated insulating states and unconventional superconductivity in twisted bilayer graphene ~\cite{Cao1, Cao2,tbg1,tbg2, koshino1, koshino2, chebrolu,balents}.  These systems are well-described in terms of (effectively) periodic superlattices with large unit cells, which under appropriate circumstances support nearly flat bands \cite{Bistritzer_2011}.  For this reason they are intriguing platforms for electron correlation physics~\cite{tbgcor1,tbgcor2,tbgcor3,tbgcor4,tbgcor5,tbgcor6,tbgcor7,tbgcor8,tbgcor10,tbgcor11,tbgcor12,tbgcora,tbgcorb},.  Beyond this,
the fragile topology of the bands \cite{Po_2019}, as well as degeneracies due to valley and spin degrees of freedom, allow competition as well as coexistence among correlated phases and broken symmetries, where quantum geometric quantities can play a decisive role.  An example of such a system is
the twisted double bilayer graphene (TDBG) system, a pair of Bernal stacked bilayer graphene systems placed in close proximity with a twist angle between them. This system may also host relatively flat moir{\'e} bands for a substantial range of twist angles, but, additionally, their Chern numbers can be tuned by application of a perpendicular electric field. The interplay and coexistence of topological order along with strong correlation effects make the system particularly intriguing~\cite{tdbgcr1,tdbgcr2, tdbgcr3, tdbgcr4, tdbgcr5, Adak_2022,mandar2, mandar3,vafek18,lado}.

\begin{figure*}
	\centering
	\includegraphics[width=\textwidth]{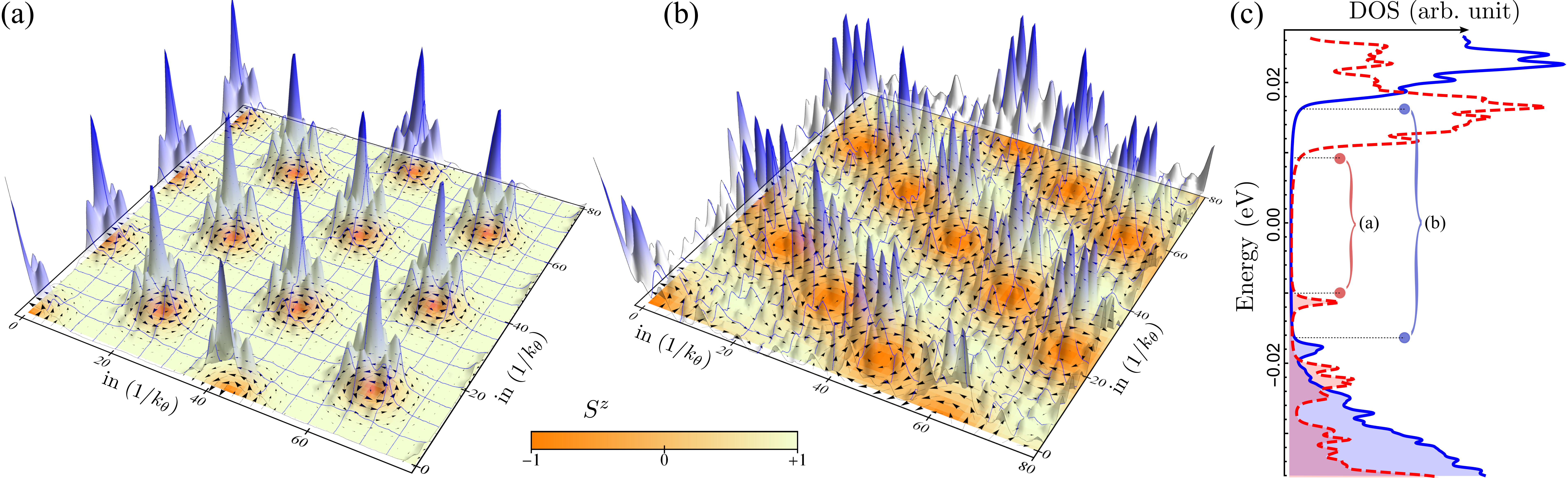}
\caption{Skyrme crystals with a supercell containing (a) 36 and (b) 72 \moire unit cells. Spin-densities of the states are plotted in the 2D plane, with excess charge density profile shown as a superimposed 3D plot. For spin-densities, the color bar represents the $z$ component of the spin.  In-plane components are drawn as vectors. Panel (a) illustrates a state with a single skyrmion per supercell, whereas (b) illustrates as state with two, so that the two cases are at the same electron density.  The latter results in a striped configuration and is energetically more favorable. (c) HF single particle density of states (DOS) for the two cases. A larger energy gap across the Fermi energy is obtained for the striped configuration, suggesting a lower energy state.}\label{fig:skyrme}
\end{figure*}

In the flat-band limit, the ground state of TDBG when the first conduction band is half-filled is predicted to be a spin-polarized insulator \cite{dassarma}.  This is experimentally supported by an observed insulating gap that increases with parallel magnetic field, presumably due to Zeeman coupling \cite{tdbgferro1,tdbgferro2,tdbgferro3,tdbgferro4,He_2020}. An interesting question is how the state evolves as the system is doped away from this particular filling.
This is the subject of our study. The simplest possible resulting state
would form a lattice of localized spin majority holes or spin minority electrons \cite{Padhi_2018,Zhang_2021,Padhi_2021,Li_2021}, in which a competition of Coulomb repulsion and the moir{\'e} lattice potential determines the lattice structure.  However, the non-trivial band topology suggests the system can further lower its energy by nucleating topological solitons in the spin configuration, which incorporate the added or removed charge in the form of skyrmions \cite{Skyrmion_book}.  Doped quantum Hall ferromagnets are believed to host skyrmions under some circumstances because the underlying Landau levels hosting them are energetically flat and carry unit Chern number \cite{Brey_1995,Perspectives_book,Goerbig21}.  Because of the combination of lattice potential and non-trivial distribution of Berry flux in a TDBG band, as we demonstrate below, TDBG ferromagnets form states which are unique to this system:   (i) Charges added or removed from the integrally filled ferromagnetic state accumulate in a single valley, leaving the one of the two valleys fully filled and spin-polarized while the other is partially filled.   (ii) The ground state forms {\it skyrmion stripes} which break the $C_3$ symmetry of the lattice. Broken orientational symmetry can be seen in both the spin and charge of the doped ground state.  This pattern of symmetry-breaking should be observable as an anisotropy in charge and energy transport through the system.

Figs. \ref{fig:skyrme} (a) and (b) illustrate skyrme states without and with stripe order, respectively, as computed in a Hartree-Fock approach which we describe below. The two states have the same electron density, but the latter is energetically favorable.  In general we find that for the lowest energy states,
the skyrmions reside in a single valley while the other valley hosts a fully-filled spin-polarized band,
which we use to define the spin $\hat{z}$ direction.  The spin density along this direction is shown in the color scale, while arrows depict the density of spin components perpendicular to this.  The broken $C_3$ symmetry in both the charge and spin order is apparent in Fig. \ref{fig:skyrme}(b).

Surprisingly, the spatial orientation of the stripes in the spin order is subtly connected to the orientation of spins in the filled valley. When the spin state of Fig. \ref{fig:skyrme}(b) is plotted after a permutation of axes such that the filled valley no longer defines the out-of-plane direction (e.g., $S_x \rightarrow S_y$, $S_y \rightarrow S_z$, $S_z \rightarrow S_x$), one finds stripes in the spin configuration rotated by 120$^\circ$ in real space relative to Fig. \ref{fig:skyrme}(b).  By contrast, the charge stripes are (necessarily) insensitive to this change.  Thus the stripe orientation is not simply a function of how the spin axes are labeled.  Indeed, in the absence of intervalley coupling, no spin direction in the valley hosting the skyrmions is picked out, and stripe order in the charge is {\it lost}.  The orientational order of the skyrmion stripes thus emerges through a subtle interplay of ferromagnetism and intervalley interactions.

\section{Moir{\'e} Model and Hartree-Fock Approach}
We model the TDBG system using the continuum approach of Refs.~\onlinecite{koshino1,chebrolu}, which follows the general approach first introduced in Ref.\onlinecite{Bistritzer_2011}; details of our implementation are given in the Appendix.  We take
the twist angle to be $1.2^o$, for which one finds a relatively flat band for each of the two valleys and two spins. The precise dispersion and topological (Chern) indices of these low-energy bands can be adjusted by applying an inter-layer electric potential $V$. In our study we considered two parameter choices yielding Chern numbers $C=1$ and $C=2$. For these choices of parameters, the Berry flux density turns out to be relatively delocalized across the \moire Brillouin zone (mBZ), which we find is significant in stabilizing skyrme crystal states (see Appendix.)

We are interested in understanding the interplay of topology and interactions in this system. Towards this end, we add the Coulomb interaction to the single-particle Hamiltonian, and then project into the four lowest (single-particle) energy conduction bands, which are isolated energetically from the other bands.  We then consider states of the system at fillings near $\nu=1/2$ (where $\nu=1/4$ indicates a single filled band), and look for energetically favorable states within the Hartree-Fock (HF) approximation.   In the presence of Coulomb interactions, at $\nu=1/2$ one finds a stable ferromagnetic, valley singlet state \cite{dassarma}, in qualitative agreement with experiment \cite{tdbgferro1,tdbgferro2,tdbgferro3,He_2020}.
In order to explore the nature of the state slightly away from the half-filling, it is necessary to introduce a supercell to accommodate the charge excess/deficit relative to $\nu=1/2$.  The superlattice this defines has primitive lattice vectors $N_1\bs{a}_1$, $N_2\bs{a}_2$, where $N_{1,2}$ are integers and $\bs{a}_{1,2}$ are primitive lattice vectors of the effective \moire lattice \cite{Bistritzer_2011}.  The original mBZ is then folded into a set of small Brillouin zones whose corners are located at $\bs{g}_i$, where $i\equiv (n_1,n_2)$, are the vectors $(n_1/N_1) \bs{b}_1+(n_2/N_2) \bs{b}_2$, with $0 \le n_j \le N_j$, $j=1,2$, and ${\bs b}_{1,2}$ are the primitive reciprocal lattice vectors of the original \moire lattice.

With this construction, the central quantity one computes in the HF approach is
\begin{align}M_{\bs{g}_l,\bs{g}_i}^{s\tau,s'\tau'}(\bar{\bs{k}})=\langle c^\dagger_{\bar{\bs{k}}+\bs{g}_l,\tau, s}c_{\bar{\bs{k}}+\bs{g}_l+\bs{g}_i,\tau^\prime, s^\prime}\rangle.
\end{align}
(See Appendix for further details.) In this expression, $s,s'$ represent spin indices and $\tau,\tau'$ are valley indices. The operators $c_{\bs{q}\tau s}$ annihilate electrons from states in the $(\tau,s)$ conduction band at wavevector $\bs{q}=\bar{\bs k} + \bs{g}_j$, so that $\bs{q}$ lies in the mBZ, while $\bar{\bs{k}}$ resides in the (smaller) superlattice Brillouin zone.
Once computed, the matrix $M$ may be used to compute any single particle quantity expectation value in the (HF) ground state.  For our present purpose,
a key feature is that it can encode the kind of non-colinear spin order that is the defining property of a skyrmion.  This involves the emergence of a spin-orbit coupling at the mean-field level which is {\it not} present  in the single-particle Hamiltonian for graphene \cite{Brey_1995}.
{In particular, under $C_3$ rotations one expects ${\rm arg}[M_{{\mathcal R}_3{\bf g}_i,{\mathcal R}_3 {\bf g}_l}^{s\tau,\bar{s}\tau'}({\mathcal R}_3 \bar{\bs{k}})] \approx {\rm arg}[M_{{\bf g}_i,{\bf g}_l}^{s\tau,\bar{s}\tau'}(\bar{\bs{k}})] \pm 2\pi/3$, where $s$ and $\bar{s}$ are oppositely oriented spins and ${\mathcal R}_3$ represents a $2\pi/3$ rotation. The sign of the phase increment depends on whether the unit cell contains skyrmions or antiskyrmions, and which of $s$ and $\bar{s}$ are up/down spins.  Deviation of the increments from precisely $\pm 2\pi/3$ is expected when $C_3$ symmetry is broken, as is the case for skyrmion stripe states, but their sum after three successive rotations should be $2\pi$.}

We note that a previous study \cite{Kwan_2022} found skyrmions in twisted {\it single} layer graphene via both a non-linear sigma model (NLSM) approach and a numerical HF study of a finite size system consisting of a single supercell.  This approach did not yield the stripe order reported here.  This may reflect differences between the twisted bilayer and single layer systems, or the need for a model in which the supercell is embedded in a larger crystal structure -- as is the case in our approach -- to stabilize the broken orientational symmetry.  Finally, NLSM's offer an alternative approach to the HF analysis we use in our study \cite{Perspectives_book,Bomerich_2020,Khalaf_2021,Kwan_2022}.  While of particular use in understanding fluctuation effects, it is challenging to capture the non-uniformity of the Berry curvature and the dispersion of the (nearly) flat bands in which the electrons are embedded via NLSM's.  These turn out to be important in determining the pattern of symmetry-breaking of the ground state, and can be well-described by HF.

In addition to spin, within the continuum approach the system has valley index as a discrete degree of freedom.  If lattice scale differences between wavefunctions in the different valleys are ignored, then the interaction has full SU(4) symmetry in spin and valley, which the ground state can exploit in forming textured states \cite{Yang_2006,Cote_2007,Doucot_2008,Cote_2008,Lian_2016}.  In our analysis we incorporate a Coulomb exchange interaction (see Appendix) which breaks this symmetry.  We find that the interaction generally energetically favors filling one of the two valleys fully, with the other fractionally filled, over forming states with valley coherence ($M^{s\tau,s^{\prime}\bar{\tau}} \ne 0$). This behavior can be understood as a realization of Hund's rule.  Near $\nu=1/2$ the filled valley is fully spin polarized, so that the exchange interaction acts as an effective Zeeman field acting on electrons in the partially filled valley, breaking the spin SU(2) symmetry within individual valleys.  Remarkably, the {\it spin} polarization of the filled valley impacts the {\it charge} ordering of the partially filled valley, yielding stripe order that is apparent in the charge density.  Ground state orders when intervalley exchange interactions are ignored are discussed in the Appendix.

\section{Topology and Energetics}
The spin texture associated with skyrmions can be characterized by the Pontryagin index,
\begin{align}
	Q_{\rm sk}=\frac{1}{4\pi}\int_{\text{unit cell}} \bs{S} \cdot (\partial_x\bs{S}\times\partial_y\bs{S})d^2\bs{r},
\label{eq:Pontryagin}
\end{align}
a topological quantity that counts the number of times a unit spin vector $\bs{S}$ wraps around the Bloch sphere as one traverses a unit cell.  When the spin density is formed from electrons within a band with Chern number $C$, the electric charge $\delta Q_e$ in the unit cell relative to that of a spin-polarized filled band obeys $|\delta Q_e| = |C Q_{\rm sk}|$.  Indeed, for states in which the spin direction varies slowly in space, the Pontryagin density (integrand in Eq. \ref{eq:Pontryagin}) closely follows the charge density \cite{Perspectives_book}.   Thus for fillings in which $Q_e \ne 0$, a spin-textured state can be lower in energy than a maximally spin-polarized one because the former involves charge that is more uniform, lowering the Hartree energy.  In the present system, two types of factors balance against this.  The first is the cost of introducing spin gradients in the ground state, which involves a loss of exchange energy relative to a polarized state.  Beyond this, in a band with non-trivial dispersion and non-uniform Berry curvature, the spin texture (within the HF description) needs to be realized by admixing single particle states of relatively low energy, and from locations in the Brillouin zone where there is non-vanishing Berry curvature.  For bands with highly localized Berry curvature (as might occur near a topological transition) this means the admixed states need to be close together in momentum.  This occurs when the skyrmions are dilute in real space.  Thus a skyrme lattice becomes less stable when the band parameters change in a such a way that the Berry curvature becomes localized.

This physics can be more concretely understood by considering band structures in which a spatially periodic, non-colinear Zeeman coupling (akin to the spin of a skryme lattice) is projected onto a band with non-vanishing Chern number.  Such a Zeeman field plays the role of exchange interactions in the HF calculations.  When the Berry flux is delocalized one finds a relatively large gap in the spectrum, with unequal numbers of states above and below the gap, as determined by the band Chern number.  As parameters are changed to approach a topological transition, the Berry flux of the \moire band becomes highly localized, and one finds states ``invade'' the gap, so that the superlattice state is less stable energetically.  This suggests that a system with a different Zeeman texture, or one with no texture at all, can provide a state of lower energy at the same filling.

A concrete example of this behavior is presented in the Appendix.  For our present purpose, we use this reasoning to understand the relative energetic stability of different skyrme lattices of the \moire ferromagnet by considering their HF energy band structures.

\section{Stripe vs. Triangular Skyrme Lattice States}
We self-consistently solve for the order-parameters within the HF approximation to obtain the ground state textures at filling $\delta\nu$ above the half-filled state for a \moire band with $C=1$.  Fig. \ref{fig:skyrme}(a) illustrates an example of the simplest case, a triangular skyrme lattice, formed by taking $N_1=N_2=6$, so that $\delta\nu=1/36$. The charge density, illustrated in blue, is well-approximated by a lattice of wavepackets.  The triangular lattice yields the lowest classical Madelung energy for charged point particles \cite{Bonsall_1977}, and so provides a low Hartree energy for this state.  Moreover, the commensuration of the skyrme lattice with the underlying \moire lattice allows for a relatively low single particle energy.  A direct computation confirms that $\delta Q_e=\pm C Q_{\rm sk}e$ for this state.  However, because a triangular lattice is not bipartite, the spin configuration involves a relatively large gradient along line segments connecting nearest neighbor skyrmions \cite{Brey_1995}.  This accounts for the separation of the highest occupied SL band from the bulk of filled, negative energy bands apparent in Fig. \ref{fig:skyrme}(c).

To improve upon this spin gradient energy, one may search for a bipartite lattice HF solution which does not have the goemetric frustration above.  This can be accomplished by considering states with an even number of skyrmions in a larger unit cell.  Fig. \ref{fig:skyrme}(b) illustrates the result of such a calculation, in which $N_1=8$, $N_2=9$, and each supercell contains two extra electrons above that of the spin-polarized state, so that here too $\delta\nu=1/36$.  A na\"ive expectation would be that the HF solution forms a (slightly distorted) honeycomb lattice.  Surprisingly, one instead finds a striped configuration.  Note that nearest neighbor skyrmions in each stripe are {\it not} identical: their in-plane spins are rotated by $\pi$ relative to one another.  Along a line segment connecting nearest neighbor skyrmions, the spin undergoes only a fraction of a full $2\pi$ rotation.  Thus this configuration involves considerably less spin gradient energy than the simpler triangular lattice.  As for (a), direct calculation confirms for this state $\delta Q_e=\pm C Q_{\rm sk}e$, where here $|\delta Q_e|=|Q_{\rm sk}|=2$.  The larger gap apparent in the Fig. \ref{fig:skyrme}(c) for this case can be interpreted as being due to lower spin gradient energy in this configuration than in (a), and suggests a lower overall energy for this state.  Direct calculation of the energy per particle confirms this.

The valley degree of freedom also plays an important role in stabilizing the broken rotational symmetry.  While skyrme crystal states with equal electron populations in both valleys can be found as solutions to the HF equations if intervalley exchange is ignored, we find these are always higher in energy than states with one spin-polarized, filled valley (see Appendix for details).  Moreover, in the presence of intervalley exchange, the spin polarization of the filled valley defines a spin axis that stabilizes the stripe order; without this, we find that a given state appears to have stripes running in directions rotated by $2\pi/3$ from one another if chooses the $S_x$, $S_y$, or $S_z$ as the out-of-plane spin axis, and there is {\it no} stripe order in the charge density.  By contrast, we find in the presence of exchange with the spin-polarized valley that physically {\it different} states with stripes running along any of the three principle directions of the underlying \moire lattice can be generated as HF ground states.  Stripe order is apparent in the charge density in any of these states, as illustrated in Fig. \ref{fig:skyrme}(b).  Finally, we have investigated states for $C=2$ bands, and find skryme stripe states rather similar to those discussed above (see Appendix.)

\section{Discussion}
To our knowledge, the skyrmion stripe state is unique among textured states induced by Berry curvature in a band: one does not find such states in quantum Hall ferromagnets \cite{Perspectives_book}. {(Interestingly, striped charge density order has been recently observed in the metallic regime of partially filled flat bands of TDBG~\cite{nematic}.)} The difference is due to the imperfect flatness of the band dispersion and a non-uniformity of the curvature that is absent in Landau levels.  Its presence suggests many possibilities for interesting behaviors and experimental signatures.  For example, the broken $C_3$ symmetry suggests anisotropic heat and charge transport properties for the system. (Note that for any finite cooling rate, the system should form domains that restore the symmetry at long length scales.  This could be overcome using strain to pick out a favored orientational direction.)  At zero temperature, quantum fluctuations might unlock the spin degrees of freedom between neighboring stripes, leading to a sliding state \cite{Fertig_1999,Mukhopadhyay_2001}.  Within individual stripes, quantum fluctuations could also induce transitions in which the in-plane spins of neighboring skyrmions decouple, endowing individual skyrmions with quantized angular momenta \cite{Cote_1997}.

There are many interesting finite temperature effects to explore as well.  The breaking of $C_3$ symmetry in the ground state suggests the system wil undergo a second order thermal phase transition into the ordered state \cite{Chaikin_1995}; this might be observed as a temperature scale below which anisotropic transport sets in.  Because the system has overall SU(2) spin symmetry in the absence of a magnetic field, long-range spin order will be absent for non-zero temperatures \cite{Auerbach_1994}.  However, application of a magnetic field to pin the spin direction in the filled valley lowers the broken spin symmetry to U(1), associated with rotations of the in-plane spins of the partially filled valley.  In this case an anisotropic spin stiffness could be observable via spin wave transport at low temperature \cite{Wei_2018,Fu_2021}.  With increasing temperature, the stiffness should vanish via a Kosterlitz-Thouless transition.  It is also interesting to speculate that a sufficiently strong in-plane magnetic field could collapse the spin texture, along with the exchange interactions that favor the stripe order, to induce a structural phase transition in the quasiparticle crystal.

\begin{acknowledgments} HAF
acknowledges the support of the NSF through Grant
Nos. ECCS-1936406 and DMR-1914451; of the Vice Provost for Research of Indiana University through the Faculty Research Support Program; and
thanks the Aspen Center for Physics (NSF
Grant No. 1066293) for its hospitality. AK acknowledges support from the SERB (Govt. of India) via sanction no. CRG/2020/001803, DAE (Govt. of India ) via sanction no. 58/20/15/2019-BRNS, as well as MHRD (Govt. of India) via sanction no. SPARC/2018-2019/P538/SL.
\end{acknowledgments}


\renewcommand{\thefigure}{S\arabic{figure}}
\setcounter{figure}{0}
\appendix
\begin{widetext}
\section{Band structure of TDBG }\label{TDBG_B0}

Twisted double bilayer graphene (TDBG) comprises two Bernal stacked graphene bilayers with a small relative  twist angle $\theta$. 
The Dirac points of a graphene bilayer are located at $\bs{K}_\tau=\tau(\bs{b}_2-\bs{b}_1)/3$, with $\tau=\pm 1$ being the valley index,  $\bs{b}_1=\frac{4\pi}{\sqrt{3}a}(-\frac{\sqrt{3}}{2},-\frac{1}{2})$ and $\bs{b}_2=\frac{4\pi}{\sqrt{3}a}(\frac{\sqrt{3}}{2},-\frac{1}{2})$ the reciprocal lattice vectors of the graphene bilayer, and $a=0.246$ nm the graphene lattice constant. An effective low energy model Hamiltonian for Bernal stacked bilayer graphene near these Dirac points is given by
\begin{align}
H^\tau =\left(\begin{array}{cc}
	h^\tau_t(\bs{k}_1) & g(\bs{k}_1) \\
	g^\dagger(\bs{k}_1)& h^\tau_b(\bs{k}_1)
\end{array}\right),
	\label{eq:HAB}
\end{align}
where $h^\tau_{t/b}(\textbf{k})$ is the monolayer graphene Hamiltonian and $g(\textbf{k})$ is the inter-layer coupling. These are given by
\begin{align}
	h^\tau_{t/b}(\bs{k})=\hbar v_F~ \bs{\sigma}.\bs{k}^\tau +\frac{\Delta}{2}(\mathbb{I}\mp \sigma_z)
\end{align}
and
\begin{align}
	g(\bs{k})=\left(\begin{array}{cc}
		\hbar v_4 k_-^{\tau} & \hbar v_3 k_+^{\tau}\\
		\gamma & \hbar v_4 k_-^{\tau}
	\end{array}\right),
\end{align}
with $\bs{k}^\tau=\tau(k_x,k_y)$ and $k_{\pm}^{\tau}=\tau( k_x \pm i k_y)$. Here $\gamma$ is the coupling between the dimer sites $A2$ and $B1$ which are vertically aligned in Bernal-stacked bilayer graphene, and $\Delta$ is an on-site potential difference between sublattice sites in a single layer. The velocities $v_3$ and $v_4$ are related 
to the effect of trigonal warping
and electron-hole asymmetry, respectively. The various parameter values we use are
$v_F=0.844\times 10^6 ~\rm{m/s}$, $v_3=-0.091\times 10^6 ~\rm{m/s}$, $v_4=-0.045\times 10^6 ~\rm{m/s}$, $\gamma=0.4~ \rm{eV}$, $\Delta=0.05~ \rm{eV}$\cite{dassarma,koshino1}.

\begin{figure}[t]
	\centering
	\includegraphics[width=0.5\textwidth]{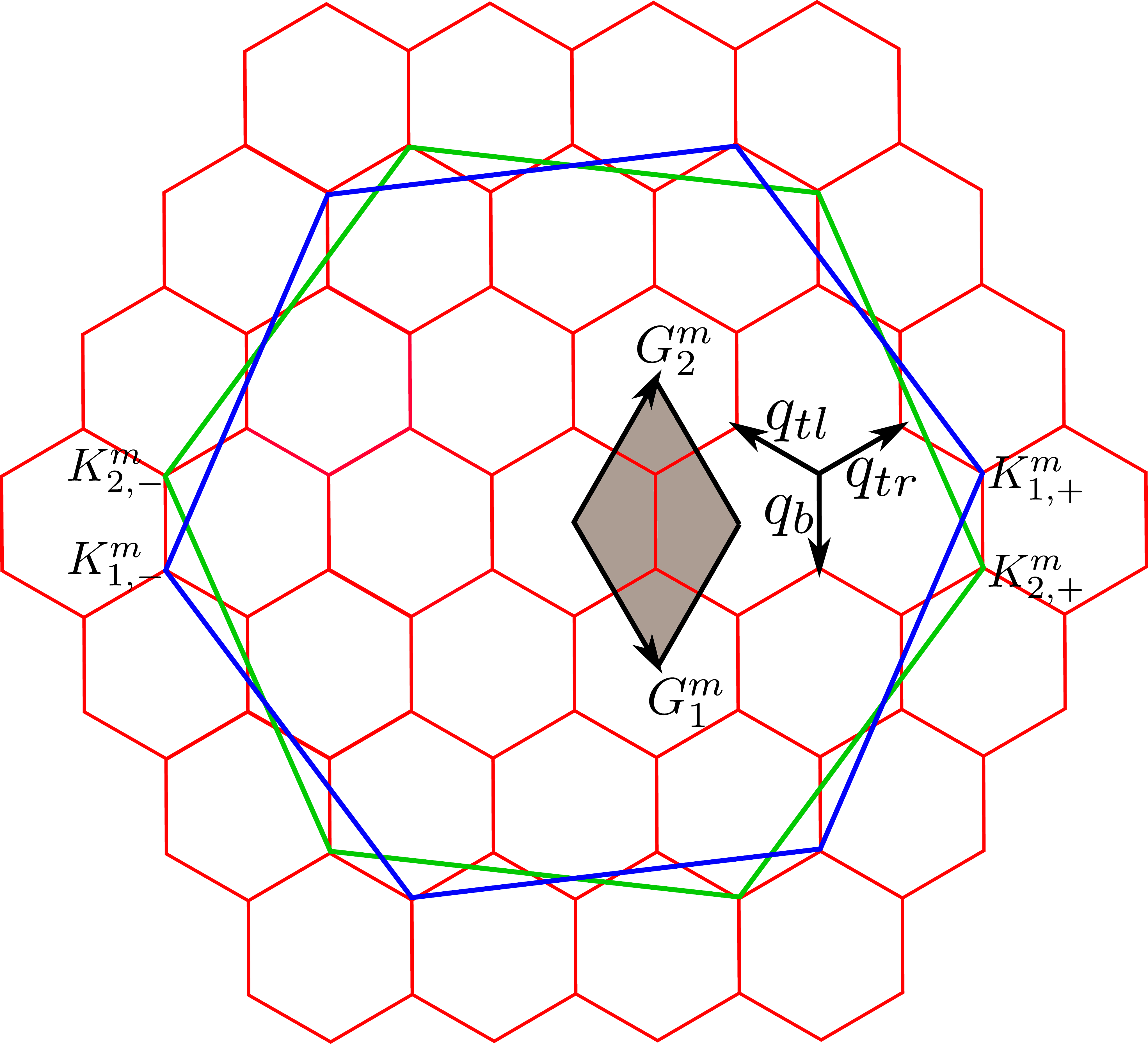}
	\caption{ \Moire structure in momentum space. The large hexagons (blue and green) represent the rotated BZ of the (top and bottom) bilayer graphene, with the top (bottom) layer rotated by $\mathcal{R}_{\theta/2}$ ($\mathcal{R}_{-\theta/2}$).  Small red hexagons represent \moire BZ's. The gray shaded rhombus represents the \moire unit cell in reciprocal space that we use in our numerics.} 
\label{lattice}
\end{figure}

Two bilayers with a small twist angle between them form an effective \moire lattice with  reciprocal lattice vector given by
\begin{align}
	\bs{G}_{1(2)}^m=(\mathcal{R}_{\theta/2}-\mathcal{R}_{-\theta/2})\bs{b}_{1(2)},
\end{align}
where $\mathcal{R}_{\theta}$ rotates a vector by an angle $\theta$ counterclockwise.  These vectors explicitly have the form $\bs{G}_{1}^m=k_\theta(\sqrt{3}/2,-3/2)$ and $\bs{G}_{2}^m=k_\theta(\sqrt{3}/2,3/2)$, with $$k_\theta=\frac{8\pi\sin{\theta/2}}{3 a}$$
the distance between nearby Dirac points of bilayer 1 and bilayer 2. The Dirac points of the \moire Brillouin zone (BZ) are located at $\bs{K}_{1(2),\tau}^{m}=\mathcal{R}_{+(-) \theta/2}\bs{K}_\tau$.  Tunneling between the two bilayers is governed by three momenta $\bs{q}_b$, $\bs{q}_{tr}$ and $\bs{q}_{tl}$ which connect the neighbouring Dirac points of the rotated bilayers.   A schematic of the \moire pattern in momentum space and the vectors described above is shown in Fig. \ref{lattice}.

There are two basic configurations for TDBG at low twist angles: AB-AB stacking and AB-BA stacking~\cite{koshino1,koshino2}. The AB-AB stacking has a $C_{3z}$ symmetry along with a $C_{2x}$ rotational symmetry, whereas the AB-BA stacking has $C_{3z}$ and $C_{2y}$ rotational symmetry. The valley Chern numbers of AB-AB and AB-BA twisted bilayer graphene differ significantly. For example, in the absence of any perpendicular electric field, undoped AB-AB twisted bilayer graphene is a trivial insulator, while the corresponding AB-BA system is a valley Hall insulator. Similarly, for a finite perpendicular electric field, the energy bands of the AB-AB and AB-BA systems have different valley Chern numbers.
The effective low energy continuum model Hamiltonian for TDBG is written in the basis of the four graphene layer sublattices (A$_1$, B$_1$, A$_2$, B$_2$, A$_3$, B$_3$, A$_4$, B$_4$ ) as
\begin{align}
	H^\tau_{\rm{AB-AB}} = \left(\begin{array}{cccc}
		h^\tau_{t}(\bs{k}_1,\theta/2) & g(\bs{k}_1,\theta/2) & 0 & 0\\
		g^\dagger(\bs{k}_1,\theta/2)& h^\tau_{b}(\bs{k}_1,\theta/2) & T(\bs{r})& 0\\
		0 & T^{\dagger}(\bs{r}) & h^\tau_{t}(\bs{k}_2,-\theta/2) & g(\bs{k}_2,-\theta/2) \\
		0 & 0 & g^{\dagger}(\bs{k}_2,-\theta/2) & h^\tau_{b}(\bs{k}_2,-\theta/2)\end{array}\right)\label{H_ABAB}
\end{align}
and
\begin{align}
	H^\tau_{\rm{AB-BA}} = \left(\begin{array}{cccc}
		h^\tau_{t}(\bs{k}_1,\theta/2) & g(\bs{k}_1,\theta/2) & 0 & 0\\
		g^\dagger(\bs{k}_1,\theta/2)& h^\tau_{b}(\bs{k}_1,\theta/2) & T(\bs{r})& 0\\
		0 & T^{\dagger}(\bs{r}) & h^\tau_{b}(\bs{k}_2,-\theta/2) & g^\dagger(\bs{k}_2,-\theta/2) \\
		0 & 0 & g(\bs{k}_2,-\theta/2) & h^\tau_{t}(\bs{k}_2,-\theta/2)\end{array}\right),\label{H_ABBA}
\end{align}
where $h^\tau_{t(b)}(\bs{k},\pm \theta/2)=h^\tau_{t(b)}(\mc{R}_{\pm \theta/2} \bs{k})$ and $g(\bs{k},\pm \theta/2)=g(\mc{R}_{\pm \theta/2}\bs{k})$. The tunneling between the two bilayers is represented by $T(\bs{r})$, which couples the bottom layer of one bilayer and the top layer of the other bilayer.  Explicitly this has the form
\begin{align}
	T(\bs{r})=T_{\tau,q_b}+e^{i\tau \bs{G}_1^m.\bs{r}} T_{\tau,q_{tl}} +e^{-i\tau \bs{G}_2^m.\bs{r}} T_{\tau,q_{tr}},
	\label{eq:T}	
\end{align}
with
\begin{align}
	T_{\tau,q_b}=\left(\begin{array}{cc}
		w_{AA} & w_{AB}\\
		w_{AB} & w_{AA}
	\end{array}\right),~~~~~~~T_{\tau,q_{tl}}=\left(\begin{array}{cc}
		w_{AA} & e^{i\tau\phi}w_{AB}\\
		w_{AB}e^{-i\tau\phi} & w_{AA}
	\end{array}\right),~~~~~T_{\tau,q_{tr}}=\left(\begin{array}{cc}
		w_{AA} & e^{-i\tau\phi}w_{AB}\\
		w_{AB}e^{i\tau\phi} & w_{AA}
	\end{array}\right).
\end{align}
Here $\phi=e^{i 2\pi/3}$ and $w_{AA}=0.05~\rm{eV}$, $w_{AB}=0.0975 ~\rm{eV}$. The value of $w_{AB}$ here is identical to that of Ref. \onlinecite{koshino1}, while the our choice of $w_{AA}$ is somewhat smaller ($w_{AA}=0.0797~\rm{eV}$ in Ref. \onlinecite{koshino1}).
Our choice of $w_{AA}$ causes the Chern band we are focusing on (band labeled $\rm{c}_1$ in Fig. \ref{fig:band} (a),(d)) to become comparatively flatter and energetically fully separated from the other bands.  This simplifies the Hartree-Fock calculation by allowing us to project away bands energetically above $\rm{c}_1$. Finally, to $H^\tau_{\rm{AB-AB}}$ or $H^\tau_{\rm{AB-BA}}$ we add a term that models the effect of
a perpendicular electric field, which has the form
\begin{align}
	\bs{V}=\left(\begin{array}{cccc}
		\frac{3V}{2}I_2 & 0 & 0 & 0\\
		0& \frac{V}{2}I_2 &0 &0\\
		0&0 & -\frac{V}{2}I_2 &0  \\
		0&0 &0 & -\frac{3V}{2}I_2
	\end{array}\right),
\end{align}
where $V$ is the applied electrostatic potential and $I_2$ is the $2 \times 2$ identity matrix.

\begin{figure*}
	\centering
	\includegraphics[width=0.95\textwidth]{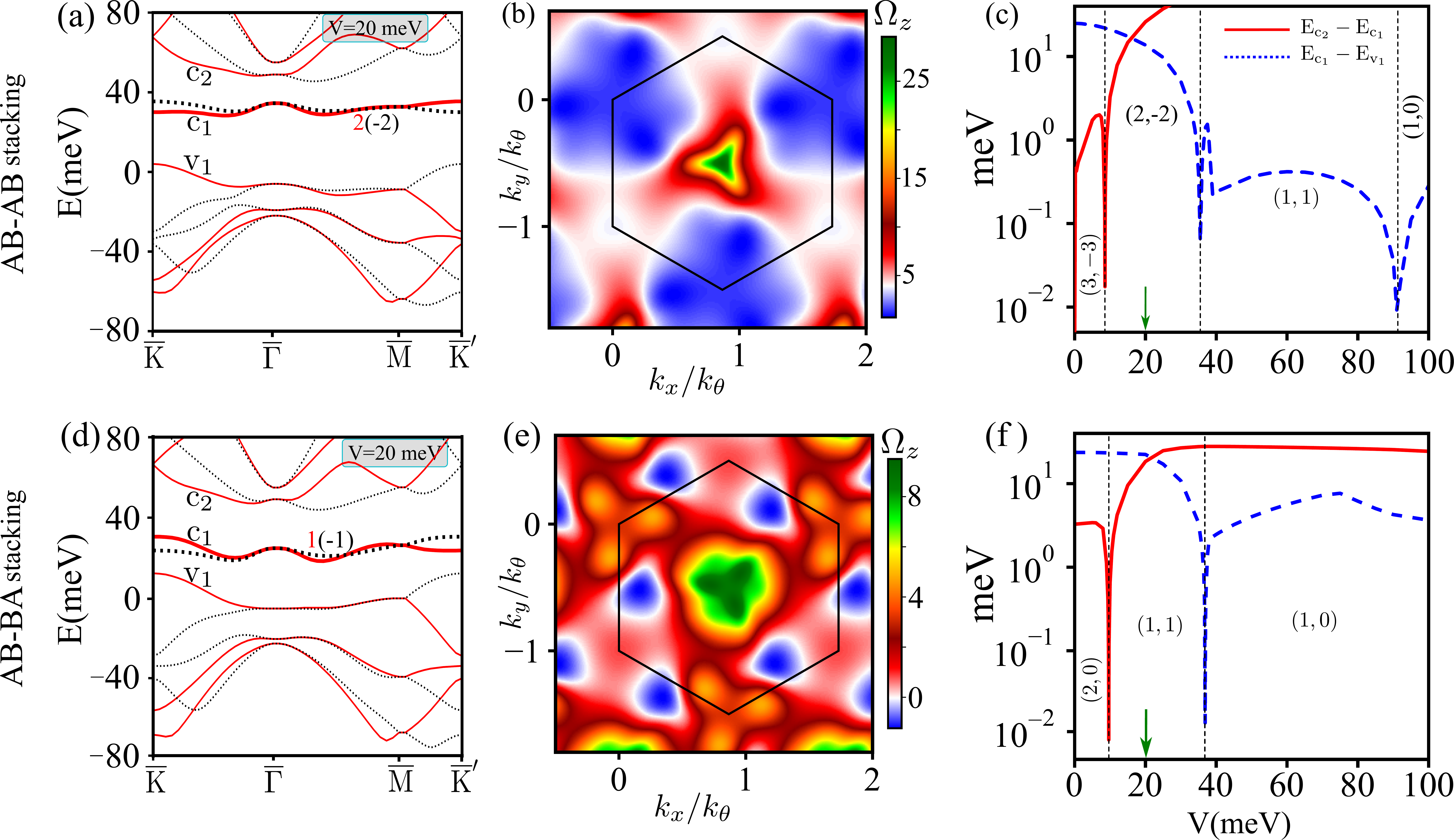}
	\caption{(a),(d): Band structure of the twisted double bilayer
		at the twist angle $\theta=1.2^o$ at V=20 \rm{meV}(perpendicular electric potential), with $w_{AA}=50$ meV, $w_{AB}=97.5$ meV. The band structure for the $\bs{K}_+$ ($\bs{K}_-$) valley is shown in red (black). The Chern number of the first \moire conduction($\rm{c}_1$) band is shown for both  stackings. (b),(e): Berry curvature of the first \moire conduction band ($\rm{c}_1$) for $\bs{K}_+$ valley. The hexagonal region represents the \moire Brillouin zone(mBZ). (c),(f): The solid red line represents the minimum energy gap between the bands labeled $\rm{c}_1$ and $\rm{c}_2$, while the blue dashed line symbolizes the energy gap between the bands labeled $\rm{c}_1$ and $\rm{v}_1$. The numbers within parentheses represent the Chern numbers of the conduction $\rm{c}_1$ and the band valence $\rm{v}_1$ band.  }\label{fig:band}
\end{figure*}

We model the TDBG system using the continuum approach of Refs. \onlinecite{koshino1,chebrolu}, which follow the general approach first introduced in Ref. \onlinecite{Bistritzer_2011}. We take
the twist angle to be $1.2^o$, for which one finds a relatively flat band for each of the two valleys and two spins, as illustrated in Figs. \ref{fig:band}(a) and (d), for AB-AB and AB-BA stackings, respectively.  The precise dispersion and topological (Chern) indices of these low-energy bands can be adjusted by applying an inter-layer electric potential $V$. Figs. \ref{fig:band}(c) and \ref{fig:band}(f) show gap closings and accompanying topological transitions that are driven by the applied potential. In our study we specifically consider two parameter choices yielding Chern numbers $C=2$ and $C=1$, at the values of $V$ indicated by arrows in Figs. \ref{fig:band}(c) and \ref{fig:band}(f), respectively. Figs. \ref{fig:band}(b) and \ref{fig:band}(d) illustrate
the Berry flux density for each of these cases, which turn out to be relatively delocalized across the \moire Brillouin zone (mBZ).  As we discuss below, this plays a significant role in stabilizing skyrmion stripe and crystal states.

\section{Non-Interacting Band structure with spin texture and role of Berry Curvature}
It is instructive to understand how a spin-texture can generate a low-energy Hartree-Fock state when bands are doped away from integral filling.  Towards this end we consider the problem of non-interacting electrons moving in a non-colinear Zeeman field, with the field carrying the same topology as a skyrme lattice.  This field plays the role of a local exchange potential that arises when spin a texture is present.
The effective Zeeman with texture we analyze has the form
\begin{align}
	h^z_1(\bs{r})&=\frac{3}{2}(\sin{\bs{g}_2.\bs{r}}-\sin{\bs{g}_1.\bs{r}})\sigma_x -\frac{\sqrt{3}}{2}(\sin{\bs{g}_1.\bs{r}}+\sin{\bs{g}_2.\bs{r}}-2\sin{\bs{g}_3.\bs{r}})\sigma_y\nonumber \\
	&+2(\cos{\bs{g}_1.\bs{r}}+\cos{\bs{g}_2.\bs{r}}+\cos{\bs{g}_3.\bs{r}})\sigma_z,\label{tex1}
\end{align}
where $\bs{g}_1=\bs{G}_1^m/N_1$,~~$\bs{g}_2=\bs{G}_2^m/N_2$,~~$\bs{g}_3=-\bs{g}_1-\bs{g}_2$ and $\sigma_x$, $\sigma_y$, and $\sigma_z$ are Pauli matrices acting on the spin degrees of freedom. This non-colinear Zeeman field produces a textured Zeeman field in real space that is periodic in a supercell containing $N_1\times N_2$ unit cells of the original \moire lattice.  The field texture is equivalent to having one skyrmion per supercell. If we project the Hamiltonian of Eq. (\ref{tex1} onto a Chern band with the Chern number $C$, the resulting spectrum has two groups of bands with a relatively large energy gap separating them, with $N_1N_2 + C$ bands in the lower energy group and $N_1N_2 - C$ bands in the higher one. Fully occupying the lower set of bands generates a relatively stable, low-energy state with filling $1+C/N_1N_2$.

The simultaneous presence of a spin texture and a Chern number is crucial to producing such a low-energy state at this charge density.  If we eliminate the texture, for example by projecting a periodic unidirectional Zeeman field into the Chern band, the large gap that opens has $N_1N_2$ bands both above and below, so that a state with the density of electrons requires occupation of some of the high energy minibands.  As a concrete example we consider a Zeeman Hamiltonian of the form
\begin{align}
	h^z_{\rm{uni}}=0.045(\sqrt{7}(\cos{\bs{g}_1.\textbf{r}}+\cos{\bs{g}_2.\textbf{r}}+\cos{\bs{g}_3.\bs{r}})\sigma_z + 10\sigma_z,\label{texuni})
\end{align}
projected into a Chern band and then diagonalized.  Fig. \ref{spinvar} illustrates the results for Eqs. \ref{tex1} and \ref{texuni} projected into a band with Chern number 1, using
$N_1=N_2=4$. The former naturally produces a low energy state for the original Chern band away from integer filling.

\begin{figure}
	\centering
	\includegraphics[scale=0.2]{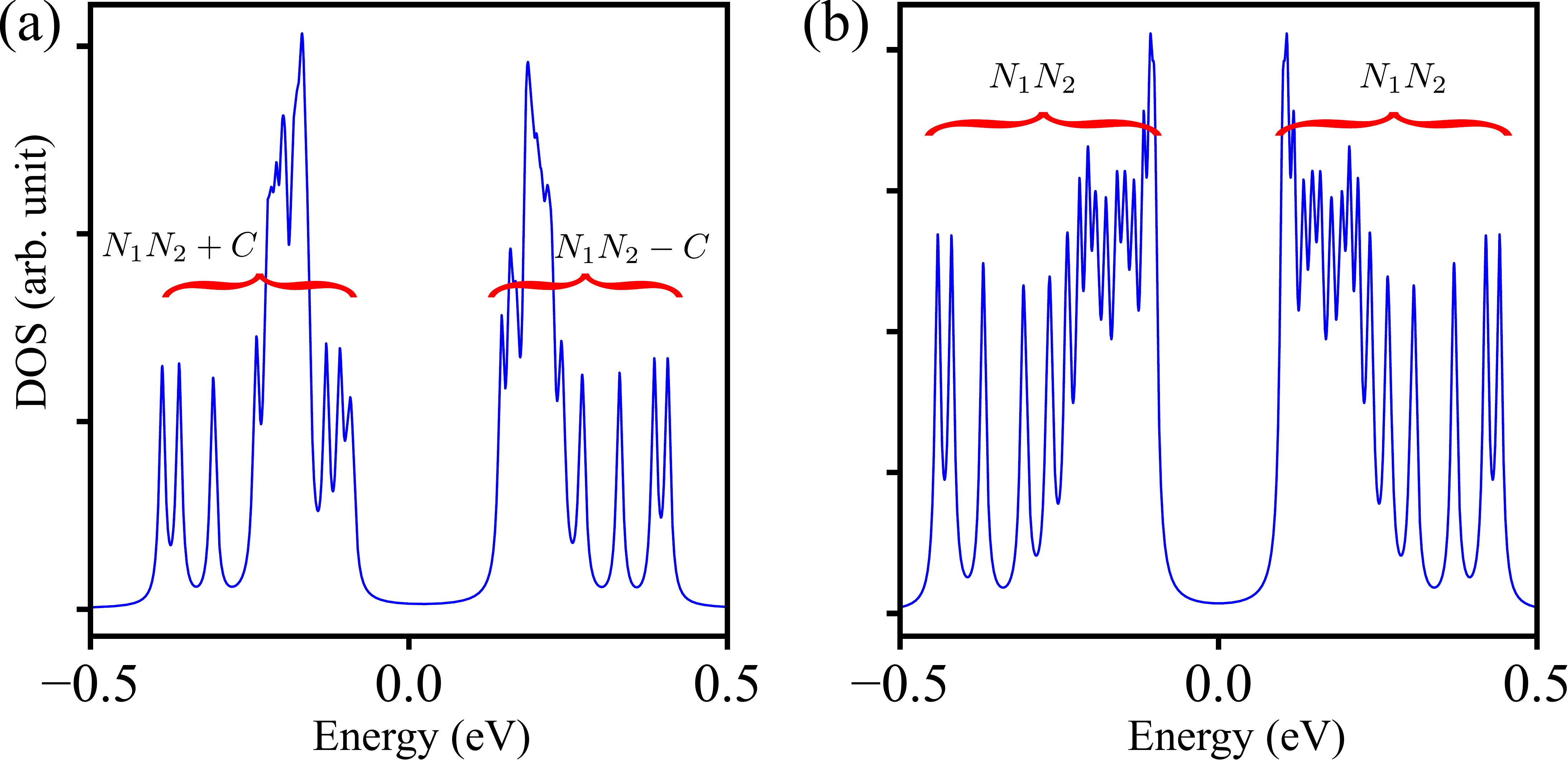}
	\caption{We project the Hamiltonian defined by Eq. (\ref{tex1}) and (\ref{texuni}) onto the Chern band labeled as $\rm{c}_1$ in Fig. \ref{fig:band}(d) for $\bs{K}_+$ valley. The resultant density of states (DOS) is shown for both the cases. (a): The energy spectrum from the projected Hamiltonian of Eq. (\ref{tex1}) opens up a gap at  $N_1N_2 \pm C$. (b): The energy spectrum from the projected Hamiltonian of Eq. (\ref{texuni}) opens up a gap at $N_1N_2$ .}\label{spinvar}
\end{figure}

The energetic stability of a doped Chern band with a spatially periodic texture at is not determined solely by the Chern number of the band.  When the Berry curvature associated with the Chern number is very localized, as occurs near a transition between Chern numbers, the enhanced stability associated described above can be undermined.  To illustrate this we evaluate the density of states for the Zeeman field in Eq. \ref{tex1} for a Chern band drawn from the TDBG spectrum at different values of electric potential $V$, such that the band changes Chern number at $V \approx 9.8$meV.  Fig. \ref{berryvar2} illustrates the Berry curvature for this series of parameters, where one can see the curvature becomes quite localized in the mBZ near the transition.  Fig. \ref{berrystandard} quantifies this degree of localization as a standard deviation $\sigma_{\Omega}$ measuring the fluctuation of the Berry curvature through the mBZ. The standard deviation ($\sigma_{\Omega}$) is defined as $\sigma_{\Omega}=\sqrt{\frac{\sum (\sigma_{\bf k} -\mu)^2 }{N}}$, where $\sigma_{\bf k}$ represents the Berry curvature at a specific momentum grid point ${\bf k}$  within the mBZ, $\mu$ denotes the average Berry curvature across the mBZ and $N$ stands for the total number of ${\bf k}$ grid points we retain within the mBZ.     In the region of large $\sigma_{\Omega}$ one sees that a group of states invade the gap (Fig. \ref{bandV}), as there is spectral transfer between the upper and lower groupings of bands.  Although the Chern number of the band is well-defined except precisely at the transition point, near the transition, where $\sigma_{\Omega}$ is large, the gap between filled and empty states is necessarily smaller than is possible for bands with smaller $\sigma_{\Omega}$.  For indicates that any lowering of energy from introducing a texture in a Chern band doped away from integral filling will be smaller for bands with larger $\sigma_{\Omega}$.

\begin{figure}
	\centering
	\includegraphics[width=0.9\textwidth]{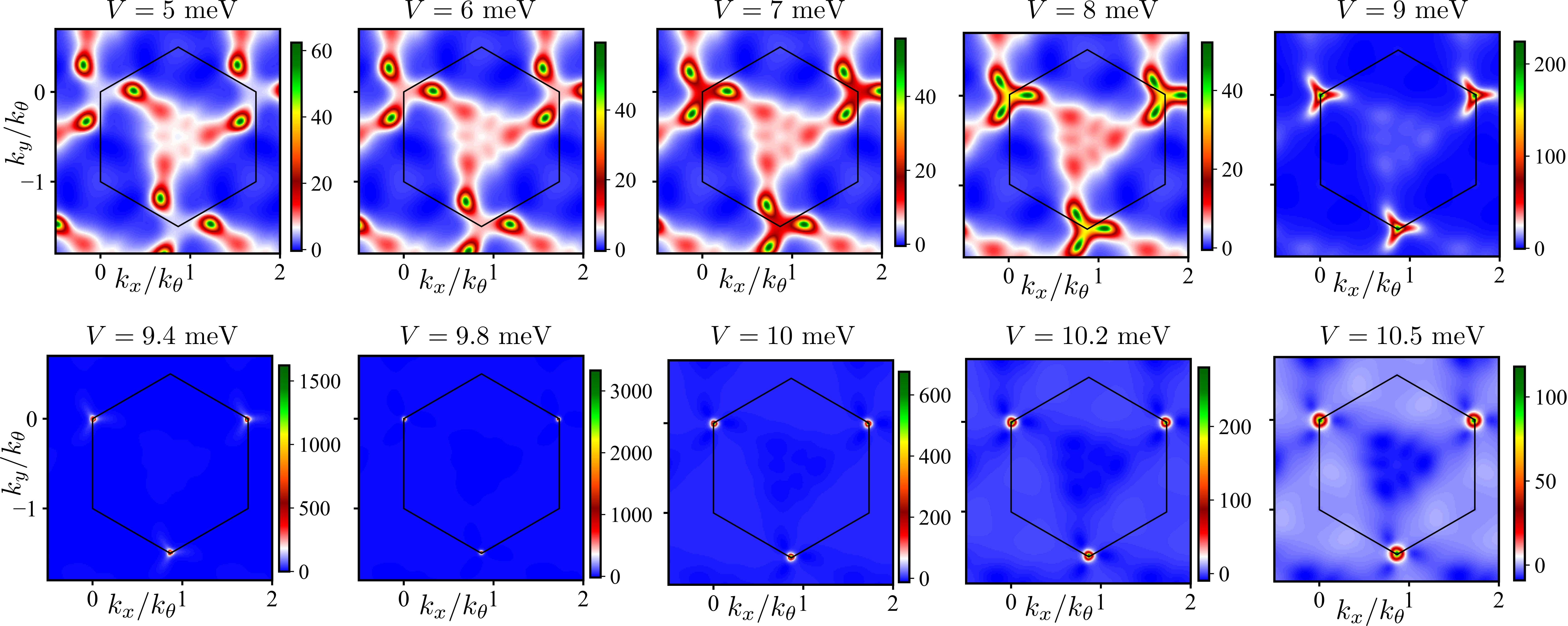}
	\caption{Variation of Berry curvature as a function of electric potential, $V$. The Berry curvature is calculated for the Chern band labeled as $\rm{c}_1$ in Fig. \ref{fig:band}(d) for $\bs{K}_+$ valley. The hexagonal region serves as the \moire Brillouin zone (mBZ). As one changes the perpendicular electric potential, the Berry curvature distribution in the mBZ is changed continuously. Around the electric potential of $V \approx 9.8~\rm{meV}$, a topological phase transition occurs, resulting in the significant localization of the Berry curvature around the Dirac points. The topological phase transition drives the Chern band with Chern number, $C=2$ to $C=1$. 
	}\label{berryvar2}
\end{figure}
~~~~~~~~~~~~~~~~~~~~

\begin{figure}
	\centering
	\includegraphics[width=0.4\textwidth]{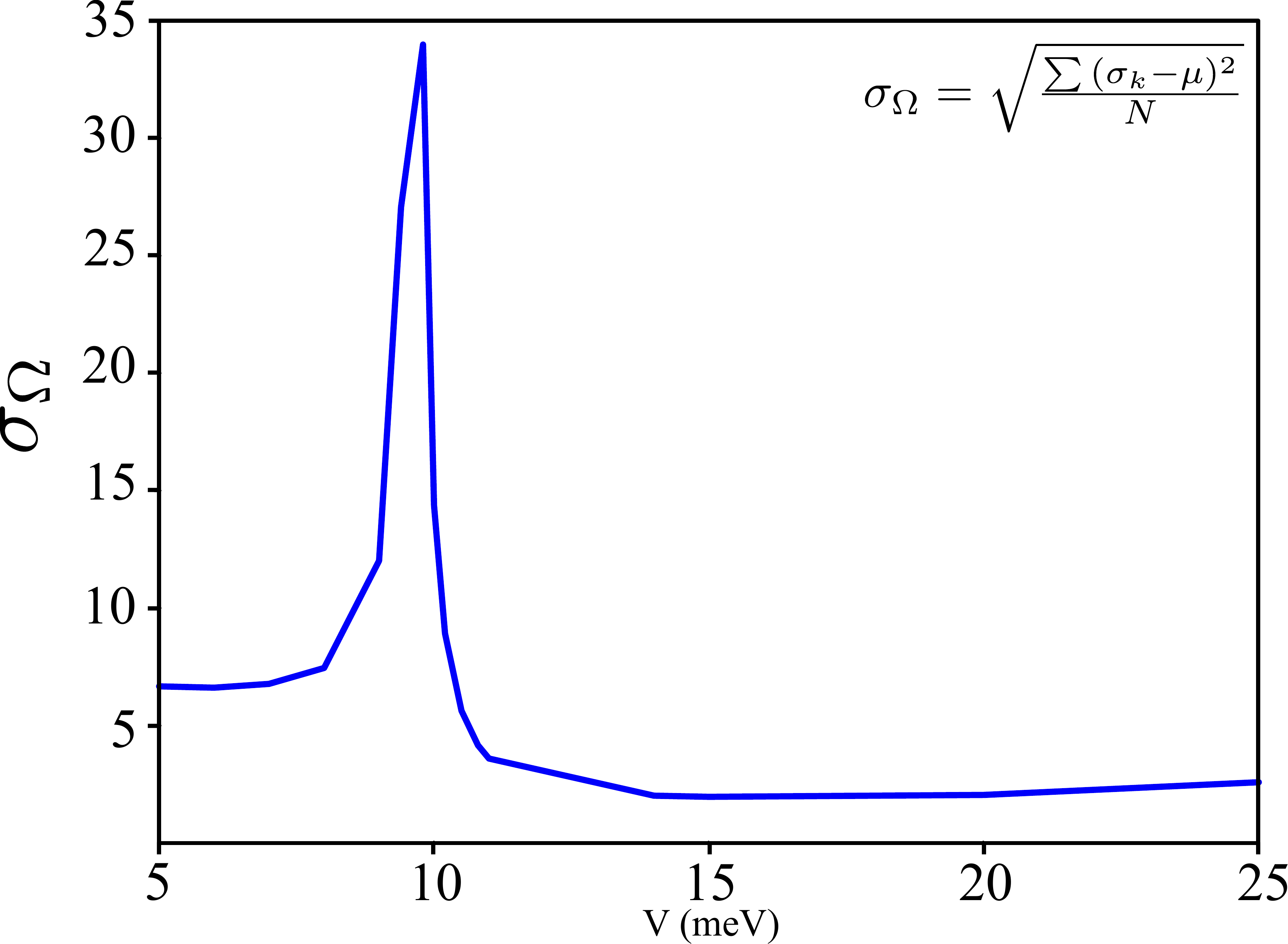}
	\caption{Standard deviation ($\sigma_{\Omega}$) of Berry curvature as a function of electric potential for the Chern band labeled $\rm{c}_1$ in Fig. \ref{fig:band}(d) for $\bs{K}_+$ valley. The Berry curvature becomes highly localized near the topological phase transition ($V\approx 9.8 \rm{meV}$), resulting in a correspondingly high standard deviation.}\label{berrystandard}
\end{figure}

\begin{figure}
	\centering
	\includegraphics[width=1.0\textwidth]{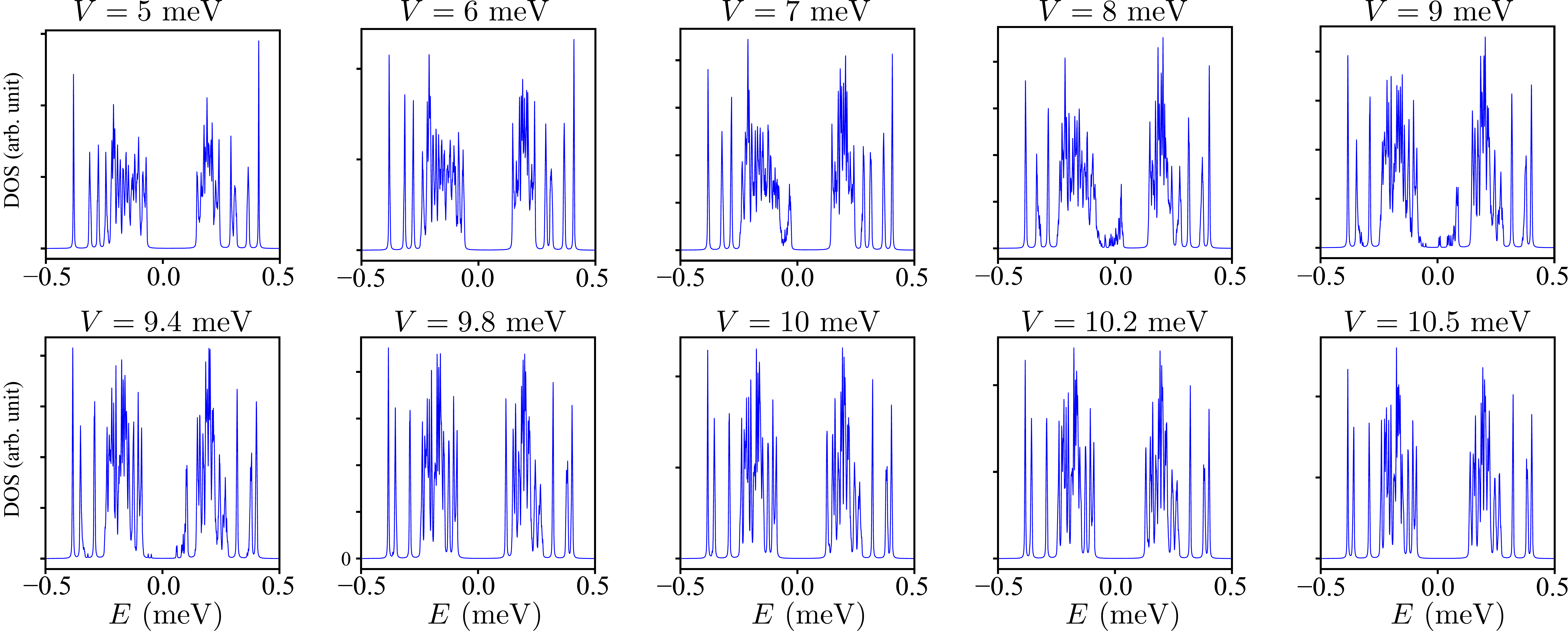}
	\caption{Density of states (DOS) derived from the energy spectrum at various values of the perpendicular electric potential.  The
		Hamiltonian (Eq. \ref{tex1}) has been projected onto the Chern band labeled $\rm{c}_1$ in Fig. \ref{fig:band}(d) for the $\bs{K}_+$ valley. Near the topological phase transition, a spectral transfer occurs between the higher and lower sets of bands.
	}\label{bandV}
\end{figure}

\section{Interaction Hamiltonian \& Hartree-Fock Approximation}
In this section we provide details of our Hartree-Fock calculation, retaining only the TDBG first conduction band ${\rm c}_1$ (see Fig.~\ref{fig:band}) of the non-interacting Hamiltonian. We start by writing the effective single particle Hamiltonian keeping only this band as
\begin{align}
	\rm{H}_0=\sum_{\bs{k},\tau,s} \xi_{\bs{k},\tau}~ c^{\dagger}_{\bs{k},\tau,s}~ c_{\bs{k},\tau,s},
\end{align}
where $c_{\bs{k},\tau,s}$ is the electron annihilation operator of the ${\rm c}_1$ band with momentum (in the \moire Brillouin zone) $\bs{k}$, in the valley $\tau=\pm 1$ and spin with $s$. $\xi_{\bs{k},\tau}$ is the spin-degenerate energy. We then project the Coulomb interaction onto the band. The interaction Hamiltonian has the form
\begin{align}
	\rm{H}_{int}&=\frac{1}{2}\int d\bs{r}_1 \int d\bs{r}_2~\rho(\bs{r}_1) V(|\bs{r}_1-\bs{r}_2|) \rho(\bs{r}_2),
	\label{eq:hint1}
\end{align}
where $V(|\bs{r}-\bs{r}^{\prime}|)$ is the screened Coulomb potential (details provided below). 
The density operator $\rho(\bs{r})$ is given by
\begin{align}
	\rho(\bs{r})&=\Psi^\dagger(\bs{r})\Psi(\bs{r})
\end{align}
where the field operator is $\Psi(\bs{r})=\sum_{\bs{k},\tau,s} \phi_{\bs{k},\tau,s}(\bs{r}) c_{\bs{k},\tau,s}$.
Here $\phi_{\bs{k},\tau,s}(\bs{r})$  are the Bloch wavefunctions of the TDBG band ($\bs{k}$ belonging to the mBZ).  These are written in terms of the periodic parts of the wavefunctions, $u_{\bs{k},\tau,s}(\bs{r})$, as
\begin{align}
	\phi_{\bs{k},\tau,s}(\bs{r})=e^{i\bs{K}_\tau.\bs{r}}e^{i\bs{k}.\bs{r}} u_{\bs{k},\tau}(\bs{r})= \frac{1}{\sqrt{\Omega}}\sum_{\bs{G}} e^{i(\bs{G}+\bs{k}+\bs{K}_\tau).\bs{r}}u_{\bs{k},\tau}(\bs{G}), \label{eq:bwf1}
\end{align}
where $\bs{K}_\tau$ are the Dirac nodes of original graphene lattice, $\bs{G}=n\bs{G}_1^m +n^\prime \bs{G}_2^m$ are the reciprocal lattice vectors of \moire lattice with $n,n^\prime \in \mathbb{Z}$. $\bs{k}$ lies within the \moire BZ and $\Omega$ is the area of a \moire unit cell. Here $u_{\bs{k},\tau}(\bs{G})$ is normalized such that
\begin{align}
	\sum_{\bs{G}} u^{\dagger}_{\bs{k},\tau}(\bs{G})
	u_{\bs{k},\tau}(\bs{G})=1.
\end{align}
Additionally we choose a gauge which satisfies
\begin{align}
	u_{\bs{k}+\bs{G}_0}(\bs{G})=u_{\bs{k}}(\bs{G}+\bs{G}_0)\label{eq:gauge}.
\end{align}	
The density operator can then be written as
\begin{align}
	\rho(\bs{r})=\Psi^\dagger(\bs{r})\Psi(\bs{r})
	=\frac{1}{\Omega}\sum_{s,\tau_1,\tau_2} \sum_{\bs{k}_1,\bs{q},\bs{G}} e^{i(\bs{K}_{\tau_2}-\bs{K}_{\tau_1}).\bs{r}} e^{i(\bs{q}+\bs{G}).\bs{r}}~ \lambda^{\tau_1,\tau_2}_{\bs{G}}(\bs{k}_1,\bs{q}) ~c^\dagger_{\bs{k}_1,\tau_1,s}c_{\bs{k}_1+\bs{q},\tau_2,s},\label{eq:density}
\end{align}
where
\begin{align}
	\lambda^{\tau_1,\tau_2}_{\bs{G}}(\bs{k}_1,\bs{q})=\sum_{\bs{G}_1}u^\dagger_{\bs{k}_1,\tau_1}(\bs{G}_1)u_{\bs{k}_1+\bs{q},\tau_2}(\bs{G}_1+\bs{G})\label{eq:lambda}
\end{align}
is the form factor. If $\bs{k}_1+\bs{q}$ lies outside the mBZ, we bring it back to the mBZ using the gauge choice of Eq.~(\ref{eq:gauge}). The interacting Hamiltonian in Eq.
(\ref{eq:hint1}) can now be written as
\begin{align}
	H_{\rm{int}}&=\frac{1}{2}\int d\bs{r}_1 \int d\bs{r}_2~\rho(\bs{r}_1) V(|\bs{r}_1-\bs{r}_2|) \rho(\bs{r}_2) \nonumber\\
	&=\frac{1}{2\Omega^2}\int d\bs{r}_1 \int d\bs{r}_2 V(|\bs{r}_1-\bs{r}_2|) \sum_{s,s^\prime,\tau_1,\tau_2,\tau_3,\tau_4} \sum_{\bs{k}_1,\bs{k}_2,\bs{q},\bs{q}^\prime,\bs{G},\bs{G}^\prime} e^{i(\bs{K}_{\tau_2}-\bs{K}_{\tau_1}).\bs{r}_1} e^{i(\bs{K}_{\tau_4}-\bs{K}_{\tau_3}).\bs{r}_2}\nonumber \\ &~e^{i\left(\bs{q}^\prime+\bs{G}^\prime\right).\bs{r}_1}~e^{i\left(\bs{q}+\bs{G}\right).\bs{r}_2} ~\lambda^{\tau_1,\tau_2}_{\bs{G}^\prime}(\bs{k}_1,\bs{q}^\prime)~\lambda^{\tau_3,\tau_4}_{\bs{G}}(\bs{k}_2,\bs{q}) ~c^\dagger_{\bs{k}_1,\tau_1,s}~c^\dagger_{\bs{k}_2,\tau_3,s^\prime}~c_{\bs{k}_2+\bs{q},\tau_4,s^\prime}~c_{\bs{k}_1+\bs{q}^\prime,\tau_2,s} \label{eq:hint}.
\end{align}

$H_{\rm int}$ contains terms that involve densities formed from field operators in the same valley, and terms where the field operators are in different valleys.  In Eq. \ref{eq:hint}, the former satisfy $\tau_1=\tau_2$ and $\tau_3=\tau_4$, and produce a contribution to the interaction given by
\begin{align}
	H^{(1)}_{\rm{int}}&=\frac{1}{2\Omega^2}\int d\bs{r}_1 \int d\bs{r}_2 V(|\bs{r}_1-\bs{r}_2|) \sum_{s,s^\prime,\tau,\tau^\prime} \sum_{\bs{k}_1,\bs{k}_2,\bs{q},\bs{q}^\prime,\bs{G},\bs{G}^\prime} ~e^{i\left(\bs{q}^\prime+\bs{G}^\prime\right).\bs{r}_1}~e^{i\left(\bs{q}+\bs{G}\right).\bs{r}_2}\nonumber \\ &~\lambda^{\tau,\tau}_{\bs{G}^\prime}(\bs{k}_1,\bs{q}^\prime)~\lambda^{\tau^\prime,\tau^\prime}_{\bs{G}}(\bs{k}_2,\bs{q}) ~c^\dagger_{\bs{k}_1,\tau,s}~c^\dagger_{\bs{k}_2,\tau^\prime,s^\prime}~c_{\bs{k}_2+\bs{q},\tau^\prime,s^\prime}~c_{\bs{k}_1+\bs{q}^\prime,\tau,s} \nonumber \\
	&=\frac{1}{2\Omega}\sum_{s,s^\prime,\tau,\tau^\prime} \sum_{\bs{k}_1,\bs{k}_2,\bs{q},\bs{G}} V(|\bs{q}+\bs{G}|)~ [\lambda^{\tau,\tau}_{G}(\bs{k}_1-\bs{q},\bs{q})]^* \lambda^{\tau^\prime,\tau^\prime}_{\bs{G}}(\bs{k}_2,\bs{q}) ~c^\dagger_{\bs{k}_1,\tau,s}~c^\dagger_{\bs{k}_2,\tau^\prime,s^\prime}~c_{\bs{k}_2+\bs{q},\tau^\prime,s^\prime}c_{\bs{k}_1-\bs{q},\tau,s} \nonumber \\
	&=\frac{1}{2\Omega}\sum_{s,s^\prime,\tau,\tau^\prime} \sum_{\bs{k}_1,\bs{k}_2,\bs{q}} F^{\tau,\tau^\prime}_1(\bs{k}_1,\bs{k}_2,\bs{q})~c^\dagger_{\bs{k}_1,\tau,s}~c^\dagger_{\bs{k}_2,\tau^\prime,s^\prime}~c_{\bs{k}_2+\bs{q},\tau^\prime,s^\prime}~c_{\bs{k}_1-\bs{q},\tau,s},\label{eq:hint_intra}
\end{align}
with
\begin{align}
	&F^{\tau,\tau^\prime}_1(\bs{k}_1,\bs{k}_2,\bs{q})=\sum_{\bs{G}} V(|\bs{q}+\bs{G}|)~ [\lambda^{\tau,\tau}_{G}(\bs{k}_1-\bs{q},\bs{q})]^* \lambda^{\tau^\prime,\tau^\prime}_{\bs{G}}(\bs{k}_2,\bs{q})\nonumber.
\end{align}
Here $$V(\bs{q})=\frac{e^2\tanh(qd)}{2\epsilon \epsilon_0 q}$$ is the Fourier transform of the screened Coulomb potential with $\epsilon$ being the dielectric constant of the environment in which the \moire structure is embedded, and $d$ is the distance between top and bottom metallic gates assumed to be present. The remaining significant terms, which introduce a relatively weak  inter-valley coupling, satisfy $\tau_1=-\tau_2=\tau$ and $\tau_3=-\tau_4=-\tau$. Taken together, these provide a contribution to the interaction Hamiltonian of the form
\begin{align}
	H^{(2)}_{\rm{int}}&=\frac{1}{2\Omega^2}\int d\bs{r}_1 \int d\bs{r}_2 V(|\bs{r}_1-\bs{r}_2|) \sum_{s,s^\prime,\tau} \sum_{\bs{k}_1,\bs{k}_2,\bs{q},\bs{q}^\prime,\bs{G},\bs{G}^\prime} e^{-2i\bs{K}_\tau.(\bs{r}_1-\bs{r}_2)} ~e^{i\left(\bs{q}^\prime+\bs{G}^\prime\right).\bs{r}_1}~e^{i\left(\bs{q}+\bs{G}\right).\bs{r}_2}\nonumber \\ & ~\lambda^{\tau,-\tau}_{\bs{G}^\prime}(\bs{k}_1,\bs{q}^\prime)~\lambda^{-\tau,\tau}_{\bs{G}}(\bs{k}_2,\bs{q}) ~c^\dagger_{\bs{k}_1,\tau,s}~c^\dagger_{\bs{k}_2,-\tau,s^\prime}~c_{\bs{k}_2+\bs{q},\tau,s^\prime}~c_{\bs{k}_1+\bs{q}^\prime,-\tau,s}\nonumber \\
	&=\frac{1}{2\Omega^2}\sum_{s,s^\prime,\tau} \sum_{\bs{k}_1,\bs{k}_2,\bs{q},\bs{q}^\prime,\bs{G},\bs{G}^\prime} \int d\bs{r}V(|\bs{r}|)~ e^{i(-2\bs{K}_\tau+\bs{q}^\prime+\bs{G}^\prime).\bs{r}} \int d\bs{r}_2 ~e^{i\left(\bs{q}^\prime+\bs{G}^\prime+\bs{q}+\bs{G}\right).\bs{r}_2}\nonumber \\
	&~\lambda^{\tau,-\tau}_{\bs{G}^\prime}(\bs{k}_1,\bs{q}^\prime)~\lambda^{-\tau,\tau}_{\bs{G}}(\bs{k}_2,\bs{q}) ~c^\dagger_{\bs{k}_1,\tau,s}~c^\dagger_{\bs{k}_2,-\tau,s^\prime}~c_{\bs{k}_2+\bs{q},\tau,s^\prime}~c_{\bs{k}_1+\bs{q}^\prime,-\tau,s}\nonumber \\
	&=\frac{V(|\bs{K}_+-\bs{K}_-|)}{2\Omega}  \sum_{s,s^\prime,\tau} \sum_{\bs{k}_1,\bs{k}_2,\bs{q}} F^{-\tau,\tau}_2(\bs{k}_1,\bs{k}_2,\bs{q}) ~c^\dagger_{\bs{k}_1,\tau,s}~c^\dagger_{\bs{k}_2,-\tau,s^\prime}~c_{\bs{k}_2+\bs{q},\tau,s^\prime}~c_{\bs{k}_1-\bs{q},-\tau,s},\label{eq:hint_inter}
\end{align}
with
\begin{align}
	&F^{-\tau,\tau}_2(\bs{k}_1,\bs{k}_2,\bs{q})=\sum_{G}[\lambda^{-\tau,\tau}_{\bs{G}}(\bs{k}_1-\bs{q},\bs{q})]^*~\lambda^{-\tau,\tau}_{\bs{G}}(\bs{k}_2,\bs{q}).\label{eq:Hint}
\end{align}
Since $\bs{q}+\bs{G}$ is much smaller than $2\bs{K}_\tau$ we replace the interaction in Eq. (\ref{eq:hint_inter}) by a constant potential, $V(|\bs{K}_+-\bs{K}_-|)$. The other terms with inter-valley densities are neglected because of the large momentum ($2\bs{K}_\tau$) they introduce in the form factors, which we assume to be vanishingly small.

Thus, our model for the interaction Hamiltonian has two terms,
\begin{align}
	H_{\rm{int}}&=H^{(1)}_{\rm{int}}+H^{(2)}_{\rm{int}}\nonumber\\
	&=\frac{1}{2\Omega}\sum_{s,s^\prime,\tau,\tau^\prime} \sum_{\bs{k}_1,\bs{k}_2,\bs{q}} F^{\tau,\tau^\prime}_1(\bs{k}_1,\bs{k}_2,\bs{q})~c^\dagger_{\bs{k}_1,\tau,s}~c^\dagger_{\bs{k}_2,\tau^\prime,s^\prime}c_{\bs{k}_2+\bs{q},\tau^\prime,s^\prime}~c_{\bs{k}_1-\bs{q},\tau,s} \nonumber \\
	&+\frac{V(|\bs{K}_+-\bs{K}_-|)}{2\Omega}  \sum_{s,s^\prime,\tau}\sum_{\bs{k}_1,\bs{k}_2,\bs{q}} F^{-\tau,\tau}_2(\bs{k}_1,\bs{k}_2,\bs{q}) ~c^\dagger_{\bs{k}_1,\tau,s}~c^\dagger_{\bs{k}_2,-\tau,s^\prime}~c_{\bs{k}_2+\bs{q},\tau,s^\prime}~c_{\bs{k}_1-\bs{q},-\tau,s}.\label{eq:H_t}
\end{align}
In our calculations,
all the momenta are in units of $k_{\theta}$, which is the momentum scale of the effective \moire Brillouin zone. We set the Coulomb energy scale to $V_{\rm{C}}=\frac{e^2}{\epsilon a_M}\approx 20~ \rm{meV}$, where $\epsilon=6$ and $a_M=\frac{a}{2\sin{\theta/2}}$ represents the \moire length scale in real space.  We assume a twist angle of $\theta=1.2^{\circ}$. The strength of the inter-valley interaction, $V(|\bs{K}_+-\bs{K}_-|)$, is of the order $V_{\rm{C}}\sin{\theta/2}\approx0.2 ~\rm{meV}$.

\subsection{Hartree Fock Approximation}

In our calculations we assume the state is periodic, with a supercell whose edges are formed from integer multiples of two non-colinear \moire primitive lattice vectors.  This supercell defines a reciprocal lattice Brillouin zone which is smaller than the mBZ, and which can be used to tile the area of the mBZ.  To do this, we note that
a momentum $\bs{k}$ within the mBZ can be written as $\bs{k}=\bar{\bs{k}}+\sum \bs{g}_i$, where the vectors $\bs{g}_i$ are defined as $\bs{g}_i=(n_1/N_1) \bs{G}_1^m+(n_2/N_2) \bs{G}_2^m$, with $i\equiv (n_1,n_2)$
and $n_1$ ($n_2$) runs from zero to $N_1$ ($N_2$). Thus the lattice for the spin texture has a real space unit cell which is $N_{1}\times N_{2}$ times larger than the \moire unit cell. The momentum $\bar{\bs{k}}$ belongs to the section of the mBZ enclosed by the vectors $\bs{G}_1^m/N_1$ and $\bs{G}_2^m/N_2$, which defines a unit cell of the superlattice in reciprocal space.

Our Hartree-Fock analysis focuses on an order parameter
which we define as
\begin{align}
	\langle c^\dagger_{\bar{\bs{k}}+\bs{g}_l,\tau, s}c_{\bar{\bs{k}}+\bs{g}_l+\bs{g}_i,\tau^\prime, s^\prime}\rangle=M_{\bs{g}_l,\bs{g}_i}^{s,\tau,s',\tau'}(\bar{\bs{k}}), \label{orderpara}
\end{align}
This quantity encodes broken spatial, spin, and valley symmetries, and is determined self-consistently.
%
In our study we focus on two different situations, those with partial valley polarization, and states with valley balanced.  As discussed in the main text, generally we find the former to yield lower energy states.

\subsection{Partial Valley Polarized (PVP) Skyrmion States}
As remarked above and in the main text, for the fillings of interest we generally find our lowest energy states to have a full, spin polarized band in one of the valleys, while the other valley hosts electrons in both spin states so that a spin-texture can be formed.  The latter entails fillings away from a single electron per \moire unit cell.
The form of the order parameter for such states is
\\
\begin{align}
	\rm{for}~~~ \tau=+1,~~~~~~~~~\langle c^\dagger_{\bar{\bs{k}}+\bs{g}_l,\tau,s}c_{\bar{\bs{k}}+\bs{g}_l+\bs{g}_i,\tau^\prime,s^\prime}\rangle=M^{\tau,s,s^\prime}_{\bs{g}_l,\bs{g}_i}(\bar{\bs{k}}) \delta_{\tau,\tau^\prime}\label{eq:vpk}
\end{align}
and 
\begin{align}
	\rm{for}~~~ \tau=-1,~~~~~~~~~\langle c^\dagger_{\bar{\bs{k}}_1,\tau,s}c_{\bar{\bs{k}}_2,\tau^\prime,s^\prime}\rangle= M_0~ \delta_{\bar{\bs{k}}_2,\bar{\bs{k}}_1} \delta_{\tau,\tau^\prime} \delta_{s,\uparrow} \delta_{s,\uparrow}.\label{eq:vpkp}
\end{align}
The interaction part of the Hartree-Fock  Hamiltonian is then given by
\begin{align}
	W^{\rm{PVP}}_{\rm{HF}}&=\frac{1}{\Omega}\sum F^{\tau,\tau^\prime}_1(\bs{k}_1,\bs{k}_2,\bs{q})\left[\langle c^\dagger_{\bs{k}_2,\tau^\prime,s^\prime}c_{\bs{k}_2+\bs{q},\tau^\prime,s^\prime} \rangle c^\dagger_{\bs{k}_1,\tau,s}c_{\bs{k}_1-\bs{q},\tau,s} - \langle c^\dagger_{\bs{k}_2,\tau^\prime,s^\prime}c_{\bs{k}_1-\bs{q},\tau,s} \rangle c^\dagger_{\bs{k}_1,\tau,s}c_{\bs{k}_2+\bs{q},\tau^\prime,s^\prime} \right] \nonumber \\
	&+\frac{V(|\bs{K}_+-\bs{K}_-|)}{\Omega}\sum F^{-\tau,\tau}_2(\bs{k}_1,\bs{k}_2,\bs{q})\left[ \langle c^\dagger_{\bs{k}_2,-\tau,s^\prime}c_{\bs{k}_2+\bs{q},\tau,s^\prime} \rangle c^\dagger_{\bs{k}_1,\tau,s}c_{\bs{k}_1-\bs{q},-\tau,s} -\langle c^\dagger_{\bs{k}_2,-\tau,s^\prime} c_{\bs{k}_1-\bs{q},-\tau,s} \rangle c^\dagger_{\bs{k}_1,\tau,s} c_{\bs{k}_2+\bs{q},\tau,s^\prime}  \right],\label{eq:mf_vp}
\end{align}
where the summation runs over all the repeated indices, that we do not mention for brevity for the rest of the Appendix.
Writing this in terms of $\bar{\bs{k}}$ yields
\begin{align}
	W^{\rm{PVP}}_{\rm{HF}}&=\frac{1}{\Omega}\sum F^{\tau,\tau^\prime}_1(\bar{\bs{k}}_1+\bs{g}_{l_1},\bar{\bs{k}}_2+\bs{g}_{l_2},\bs{q})\bigg[\langle c^\dagger_{\bar{\bs{k}}_2+\bs{g}_{l_2},\tau^\prime,s^\prime}c_{\bar{\bs{k}}_2+\bs{g}_{l_2}+\bs{q},\tau^\prime,s^\prime} \rangle c^\dagger_{\bar{\bs{k}}_1+\bs{g}_{l_1},\tau,s}c_{\bar{\bs{k}}_1+\bs{g}_{l_1}-\bs{q},\tau,s} \nonumber \\
	&- \langle c^\dagger_{\bar{\bs{k}}_2+\bs{g}_{l_2},\tau^\prime,s^\prime}c_{\bar{\bs{k}}_1+\bs{g}_{l_1}-\bs{q},\tau,s} \rangle c^\dagger_{\bar{\bs{k}}_1+\bs{g}_{l_1},\tau,s}c_{\bar{\bs{k}}_2+\bs{g}_{l_2}+\bs{q},\tau^\prime,s^\prime} \bigg] \nonumber \\
	&+\frac{V(|\bs{K}_+-\bs{K}_-|)}{\Omega}\sum F^{-\tau,\tau}_2(\bar{\bs{k}}_1+\bs{g}_{l_1},\bar{\bs{k}}_2+\bs{g}_{l_2},\bs{q})\bigg[ \langle c^\dagger_{\bar{\bs{k}}_2+\bs{g}_{l_2},-\tau,s^\prime}c_{\bar{\bs{k}}_2+\bs{g}_{l_2}+\bs{q},\tau,s^\prime} \rangle c^\dagger_{\bar{\bs{k}}_1+\bs{g}_{l_1},\tau,s}c_{\bar{\bs{k}}_1+\bs{g}_{l_1}-\bs{q},-\tau,s} \nonumber \\
	&-\langle c^\dagger_{\bar{\bs{k}}_2+\bs{g}_{l_2},-\tau,s^\prime} c_{\bar{\bs{k}}_1+\bs{g}_{l_1}-\bs{q},-\tau,s} \rangle c^\dagger_{\bar{\bs{k}}_1+\bs{g}_{l_1},\tau,s} c_{\bar{\bs{k}}_2+\bs{g}_{l_2}+\bs{q},\tau,s^\prime}  \bigg]. \label{eq:mf_vp1}
\end{align}
In this case it is possible to write Hartree-Fock Hamiltonians for each valley separately.  The interaction parts of these take the form
\begin{align}
	W_{\rm{HF}}^{\rm{PVP},+1}&=\frac{1}{\Omega}\sum \bigg[F^{+,+}_1(\bar{\bs{k}}_1+\bs{g}_{l_1},\bar{\bs{k}}_2+\bs{g}_{l_2},\bs{q}) ~M^{+,s^\prime,s^\prime}_{\bs{g}_{l_2},\bs{g}_{i}}(\bar{\bs{k}}_2)~ \delta _{\bs{q},\bs{g}_{i}}\nonumber \\
	&+ F^{+,-}_1(\bar{\bs{k}}_1+\bs{g}_{l_1},\bar{\bs{k}}_2+\bs{g}_{l_2},\bs{q})~ M_0 ~\delta_{s^\prime,\uparrow}~\delta_{\bs{q},0}\bigg] c^\dagger_{\bar{\bs{k}}_1+\bs{g}_{l_1},+,s}c_{\bar{\bs{k}}_1+\bs{g}_{l_1}-\bs{q},+,s}\nonumber \\
	&-\frac{1}{\Omega}\sum F^{+,+}_1(\bar{\bs{k}}_1+\bs{g}_{l_1},\bar{\bs{k}}_2+\bs{g}_{l_2},\bs{q}) M^{+,s^\prime,s}_{\bs{g}_{l_2},\bs{g}_{i}}(\bar{\bs{k}}_2)~\delta_{\bar{\bs{k}}_1+\bs{g}_{l_1}-\bs{q} - \bar{\bs{k}}_2-\bs{g}_{l_2},\bs{g}_i} ~~ c^\dagger_{\bar{\bs{k}}_1+\bs{g}_{l_1},+,s}c_{\bar{\bs{k}}_2+\bs{g}_{l_2}+\bs{q},+,s^\prime}\nonumber \\
	&-\frac{V(|\bs{K}_+-\bs{K}_-|)}{\Omega}\sum F^{-,+}_2(\bar{\bs{k}}_1+\bs{g}_{l_1},\bar{\bs{k}}_2+\bs{g}_{l_2},\bs{q}) M_0~\delta_{\bs{q},0}~ \delta_{s,\uparrow}~\delta_{s^\prime,\uparrow} ~\delta_{\bar{\bs{k}}_1+\bs{g}_{l_1}-\bs{q} , \bar{\bs{k}}_2+\bs{g}_{l_2}}~c^\dagger_{\bar{\bs{k}}_1+\bs{g}_{l_1},+,s} c_{\bar{\bs{k}}_2+\bs{g}_{l_2}+\bs{q},+,s^\prime} \nonumber \\
	\quad
	&=\frac{1}{\Omega}\sum F^{+,+}_1(\bar{\bs{k}}_1+\bs{g}_{l_1},\bar{\bs{k}}_2+\bs{g}_{l_2},\bs{g}_i)~ M^{+,s^\prime,s^\prime}_{\bs{g}_{l_2},\bs{g}_{i}}(\bar{\bs{k}}_2)~ c^\dagger_{\bar{\bs{k}}_1+\bs{g}_{l_1},+,s}c_{\bar{\bs{k}}_1+\bs{g}_{l_1}-\bs{g}_i,+,s}\nonumber \\
	&+\frac{1}{\Omega}\sum F^{+,-}_1(\bar{\bs{k}}_1+\bs{g}_{l_1},\bar{\bs{k}}_2+\bs{g}_{l_2},0) ~M_0~~\delta_{s^\prime,\uparrow}~ c^\dagger_{\bar{\bs{k}}_1+\bs{g}_{l_1},+,s}c_{\bar{\bs{k}}_1+\bs{g}_{l_1},+,s}\nonumber \\
	&-\frac{1}{\Omega}\sum F^{+,+}_1(\bar{\bs{k}}_1+\bs{g}_{l_1},\bar{\bs{k}}_2+\bs{g}_{l_2},\bar{\bs{k}}_1+\bs{g}_{l_1}-\bar{\bs{k}}_2-\bs{g}_{l_2}-\bs{g}_i)~ M^{+,s^\prime,s}_{\bs{g}_{l_2},\bs{g}_{i}}(\bar{\bs{k}}_2)~c^\dagger_{\bar{\bs{k}}_1+\bs{g}_{l_1},+,s}c_{\bar{\bs{k}}_1+\bs{g}_{l_1}-\bs{g}_i,+,s^\prime}\nonumber \\
	&-\frac{V(|\bs{K}|_+-\bs{K}_-)}{\Omega}\sum F^{-,+}_2(\bar{\bs{k}}_1+\bs{g}_{l_1},\bar{\bs{k}}_2+\bs{g}_{l_2},\bar{\bs{k}}_1+\bs{g}_{l_1}-\bar{\bs{k}}_2-\bs{g}_{l_2}) ~M_0~\delta_{s,\uparrow}\delta_{s^\prime,\uparrow}~c^\dagger_{\bar{\bs{k}}_1+\bs{g}_{l_1},+,s}c_{\bar{\bs{k}}_1+\bs{g}_{l_1},+,s^\prime}\label{eq:mf_vpp}
\end{align}
and
\begin{align}
	W^{\rm{PVP},-1}_{\rm{MF}}&=\frac{1}{\Omega}\sum F^{-,+}_1(\bar{\bs{k}}_1+\bs{g}_{l_1},\bar{\bs{k}}_2+\bs{g}_{l_2},\bs{g}_i)~ M^{+,s^\prime,s^\prime}_{\bs{g}_{l_2},\bs{g}_{i}}(\bar{\bs{k}}_2)~ c^\dagger_{\bar{\bs{k}}_1+\bs{g}_{l_1},-,s}c_{\bar{\bs{k}}_1+\bs{g}_{l_1}-\bs{g}_i,-,s}\nonumber \\
	&+\frac{1}{\Omega}\sum F^{-,-}_1(\bar{\bs{k}}_1+\bs{g}_{l_1},\bar{\bs{k}}_2+\bs{g}_{l_2},0) ~M_0~~\delta_{s^\prime,\uparrow}~ c^\dagger_{\bar{\bs{k}}_1+\bs{g}_{l_1},-,s}c_{\bar{\bs{k}}_1+\bs{g}_{l_1},-,s}\nonumber \\
	&-\frac{1}{\Omega}\sum F^{-,-}_1(\bar{\bs{k}}_1+\bs{g}_{l_1},\bar{\bs{k}}_2+\bs{g}_{l_2},\bar{\bs{k}}_1+\bs{g}_{l_1}-\bar{\bs{k}}_2-\bs{g}_{l_2})~ M_0~\delta_{s,\uparrow}\delta_{s^\prime,\uparrow}~c^\dagger_{\bar{\bs{k}}_1+\bs{g}_{l_1},-,s}c_{\bar{\bs{k}}_1+\bs{g}_{l_1},-,s^\prime}\nonumber \\
	&-\frac{V(|\bs{K}_+-\bs{K}_-|)}{\Omega}\sum F^{+,-}_2(\bar{\bs{k}}_1+\bs{g}_{l_1},\bar{\bs{k}}_2+\bs{g}_{l_2},\bar{\bs{k}}_1+\bs{g}_{l_1}-\bar{\bs{k}}_2-\bs{g}_{l_2}-\bs{g}_i) M^{+,s^\prime,s}_{\bs{g}_{l_2},\bs{g}_{i}}(\bar{\bs{k}}_2)~c^\dagger_{\bar{\bs{k}}_1+\bs{g}_{l_1},-,s}c_{\bar{\bs{k}}_1+\bs{g}_{l_1}-\bs{g}_i,-,s^\prime}\label{eq:mf_vpm}.
\end{align}
We solve Eq. (\ref{eq:mf_vpp}) and Eq. (\ref{eq:mf_vpm})  self consistently to compute the order parameters $M^{+,s,s^\prime}_{\textbf{\textit{g}}_{l_2},\textbf{\textit{g}}_{i}}(\bar{\textbf{k}}_2)$ and $M_0$.  Note that $M_0$ 
enters $W^{PVP, +1}_{HF}$ as an effective Zeeman field that tends to align spins in the partially filled valley.

\subsection{Valley Balanced (VB) Skyrmion States}
For a state with valley balanced, the order parameter retains its original form,
\begin{align}
	\langle c^\dagger_{\bar{\bs{k}}+\bs{g}_l,\tau, s}c_{\bar{\bs{k}}+\bs{g}_l+\bs{g}_i,\tau^\prime, s^\prime}\rangle=M_{\bs{g}_l,\bs{g}_i}^{s,\tau,s',\tau'}(\bar{\bs{k}}), \label{eq:op_vc}
\end{align}
and off-diagonal terms in the valley index are non-zero.  In this situation there is a single Hartree-Fock Hamiltonian for which eigenstates will have contributions from both valleys.  The Hartree-Fock approximation for the interaction part of the Hamiltonian in this case has the form
\begin{align}
	W^{\rm{VB}}_{\rm{HF}}&=\frac{1}{\Omega}\sum F^{\tau,\tau^\prime}_1(\bar{\bs{k}}_1+\bs{g}_{l_1},\bar{\bs{k}}_2+\bs{g}_{l_2},\bs{q})\bigg[\langle c^\dagger_{\bar{\bs{k}}_2+\bs{g}_{l_2},\tau^\prime,s^\prime}c_{\bar{\bs{k}}_2+\bs{g}_{l_2}+\bs{q},\tau^\prime,s^\prime} \rangle c^\dagger_{\bar{\bs{k}}_1+\bs{g}_{l_1},\tau,s}c_{\bar{\bs{k}}_1+\bs{g}_{l_1}-\bs{q},\tau,s} \nonumber \\
	&- \langle c^\dagger_{\bar{\bs{k}}_2+\bs{g}_{l_2},\tau^\prime,s^\prime}c_{\bar{\bs{k}}_1+\bs{g}_{l_1}-\bs{q},\tau,s} \rangle c^\dagger_{\bar{\bs{k}}_1+\bs{g}_{l_1},\tau,s}c_{\bar{\bs{k}}_2+\bs{g}_{l_2}+\bs{q},\tau^\prime,s^\prime} \bigg] \nonumber \\
	&+\frac{V(|2\bs{K}|)}{\Omega}\sum F^{-\tau,\tau}_2(\bar{\bs{k}}_1+\bs{g}_{l_1},\bar{\bs{k}}_2+\bs{g}_{l_2},\bs{q})\bigg[ \langle c^\dagger_{\bar{\bs{k}}_2+\bs{g}_{l_2},-\tau,s^\prime}c_{\bar{\bs{k}}_2+\bs{g}_{l_2}+\bs{q},\tau,s^\prime} \rangle c^\dagger_{\bar{\bs{k}}_1+\bs{g}_{l_1},\tau,s}c_{\bar{\bs{k}}_1+\bs{g}_{l_1}-\bs{q},-\tau,s} \nonumber \\
	&-\langle c^\dagger_{\bar{\bs{k}}_2+\bs{g}_{l_2},-\tau,s^\prime} c_{\bar{\bs{k}}_1+\bs{g}_{l_1}-\bs{q},-\tau,s} \rangle c^\dagger_{\bar{\bs{k}}_1+\bs{g}_{l_1},\tau,s} c_{\bar{\bs{k}}_2+\bs{g}_{l_2}+\bs{q},\tau,s^\prime}  \bigg]\nonumber \\
	&\equiv \frac{1}{\Omega}\sum F^{\tau,\tau^\prime}_1(\bar{\bs{k}}_1+\bs{g}_{l_1},\bar{\bs{k}}_2+\bs{g}_{l_2},\bs{g}_i)~ M^{s^\prime,\tau^\prime,s^\prime,\tau^\prime}_{\bs{g}_{l_2},\bs{g}_{i}}(\bar{\bs{k}}_2)~ c^\dagger_{\bar{\bs{k}}_1+\bs{g}_{l_1},\tau,s}c_{\bar{\bs{k}}_1+\bs{g}_{l_1}-\bs{g}_i,\tau,s}\nonumber \\
	&-\frac{1}{\Omega}\sum F^{\tau,\tau^\prime}_1(\bar{\bs{k}}_1+\bs{g}_{l_1},\bar{\bs{k}}_2+\bs{g}_{l_2},\bar{\bs{k}}_1+\bs{g}_{l_1}-\bar{\bs{k}}_2-\bs{g}_{l_2}-\bs{g}_i)~M^{s^\prime,\tau^\prime,s,\tau}_{\bs{g}_{l_2},\bs{g}_{i}}(\bar{\bs{k}}_2)~c^\dagger_{\bar{\bs{k}}_1+\bs{g}_{l_1},\tau,s}c_{\bar{\bs{k}}_1+\bs{g}_{l_1}-\bs{g}_i,\tau^\prime,s^\prime}\nonumber \\
	&+\frac{V(|\bs{K}|_+-\bs{K}_-)}{\Omega}\sum F^{-\tau,\tau}_2(\bar{\bs{k}}_1+\bs{g}_{l_1},\bar{\bs{k}}_2+\bs{g}_{l_2},\bs{g}_i)~M^{s^\prime,-\tau,s^\prime,\tau}_{\bs{g}_{l_2},\bs{g}_{i}}(\bar{\bs{k}}_2)~ c^\dagger_{\bar{\bs{k}}_1+\bs{g}_{l_1},\tau,s}c_{\bar{\bs{k}}_1+\bs{g}_{l_1}-\bs{g}_i,-\tau,s} \nonumber \\
	&-\frac{V(|\bs{K}|_+-\bs{K}_-)}{\Omega}\sum F^{-\tau,\tau}_2(\bar{\bs{k}}_1+\bs{g}_{l_1},\bar{\bs{k}}_2+\bs{g}_{l_2},\bar{\bs{k}}_1+\bs{g}_{l_1}-\bar{\bs{k}}_2-\bs{g}_{l_2}-\bs{g}_i)~M^{s^\prime,-\tau,s,-\tau}_{\bs{g}_{l_2},\bs{g}_{i}}(\bar{\bs{k}}_2)~c^\dagger_{\bar{\bs{k}}_1+\bs{g}_{l_1},\tau,s} c_{\bar{\bs{k}}_1+\bs{g}_{l_1}-\bs{g}_i,\tau,s^\prime}.
	\label{eq:mf_vc}
\end{align}
The total mean-field Hamiltonian, therefore can be written as
\begin{align}
	H_{\rm{HF}}=H_0+W_{\rm{HF}}\label{eq:totalhf},
\end{align}
where $W_{\rm{HF}}$ is given by  Eq. (\ref{eq:mf_vpp}), (\ref{eq:mf_vpm}) or Eq. (\ref{eq:mf_vc}).

\subsection{Self-Consistency Condition}
The mean-field Hamiltonian with Eq. (\ref{eq:totalhf}) can be re-written as
\begin{align}
	H_{\rm HF} = \mathcal{C}^{\dagger}\mathcal{H}\mathcal{C}\label{Hmf},
\end{align}
where $\mathcal{C}$ is the vector containing the annihilation operators with 
\begin{align}
	\mathcal{C}(\bar{\bs{k}})\equiv \left\{\mathcal{C}^{s,\tau}(\bar{\bs{k}})\right\}\equiv\left(\mathcal{C}^{\uparrow,+}(\bar{\bs{k}}), \mathcal{C}^{\downarrow,+}(\bar{\bs{k}}), \mathcal{C}^{\uparrow,-}(\bar{\bs{k}}), \mathcal{C}^{\downarrow,-}(\bar{\bs{k}})\right),
\end{align}
and
\begin{align}
	\mathcal{C}^{s,\tau}(\bar{\bs{k}})=\left(c^{s,\tau}(\bar{\bs{k}}),c^{s,\tau}(\bar{\bs{k}}+\bs{g}_{1}),....,c^{s,\tau}(\bar{\bs{k}}+\bs{g}_{i}),....,c^{s,\tau}(\bar{\bs{k}}+\bs{g}_{N})\right).
\end{align}
One can then diagonalize the mean-field Hamiltonian matrix using unitary transformation $\mathcal{U}$,
\begin{align}
	\mathcal{U}^\dagger~ \mathcal{H} ~\mathcal{U}= D,
\end{align}
where $D$ is a diagonal matrix. The operators that diagonalize the Hamiltonian are then
\begin{align}
	\mathcal{V}=\mathcal{U}^\dagger \mathcal{C},
\end{align}
which is an $4N$ component vector with $N=N_1N_2$. The matrix elements of $M$,
\begin{align}
	M^{s,\tau,s^\prime,\tau^\prime,}_{\bs{g}_{l},\bs{g}_{i}}(\bar{\bs{k}})&=\langle c^\dagger_{\bar{\bs{k}}+\bs{g}_{l},\tau,s}~c^\dagger_{\bar{\bs{k}}_1+\bs{g}_{l}+\bs{g}_{i},\tau^\prime,s^\prime} \rangle \label{eq:self}
\end{align}
can then be constructed directly from the matrix elements of $U$.
	Chemical potential, which appears in writing the above equation in terms of Fermi distribution, is chosen to ensure that the filling of the bands gives the correct density of electrons for the system. 

For a given assumed Hartree-Fock interaction, with an assumed form of $M$, $\mathcal{H}$ is diagonalized, and Eq. (\ref{eq:self}) is used to recompute $M$.  This process is iterated until the Hartree-Fock Hamiltonian ceases to change significantly.  We consider our solutions converged when the energy difference of the states between two successive iteration is less than $10^{-5}$ eV.

\subsection{Spin and Charge Densities}
After achieving a converged solution for the order parameter, we subsequently calculate the spin and charge densities of the resulting ground state. First, one needs to write the field operator (of the relevant single conduction band), in terms of individual valley and spin resolved field operators,   $\Psi(\bs{r})\equiv\sum_{\bs{k},\tau,s}\psi_{\bs{k},\tau,s}(\bs{r}) =\sum_{\bs{k},\tau,s}\phi_{\bs{k},\tau,s}(\bs{r})c_{\bs{k},\tau,s}$, where $\phi_{\bs{k},\tau,s}$ are the Bloch wave-functions of the TDBG band, with $\bs{k}$ contained in the mBZ. The density matrix is then defined in terms of the expectation values $$\rho_{\tau,\tau'}^{ss'}(\bs{r}) \approx \delta_{\tau\tau'}\rho_{\tau}^{ss'},~~ {\rm with~~}\rho_{\tau}^{ss'}=\langle \psi^{\dagger}_{\bs{k},\tau,s}(\bs{r})\psi_{\bs{k},\tau,s}(\bs{r}) \rangle,$$
where we have used our numerical observation that such expectation values in the ground state are \textit{ diagonal in the valley index}. We can then compute a local density matrix
\begin{align}
	\rho^{ss^\prime}_{\tau}(\bs{r})&\equiv\langle \psi^\dagger_{\tau,s}(\bs{r}) \psi_{\tau,s^\prime}(\bs{r})\rangle \nonumber \\
	&=\sum u^\dagger_{\bs{k}_1,\tau}(\bs{G}_1)~u_{\bs{k}_2,\tau}(\bs{G}_2)~~e^{i\left(\bs{G}_2-\bs{G}_1+\bs{k}_2-\bs{k}_1\right).\bs{r}} \langle c^\dagger_{\bs{k}_1,\tau,s} c_{\bs{k}_2,\tau,s^\prime}\rangle \nonumber \\
	&=\sum u^\dagger_{\bar{\bs{k}}_1+\bs{g}_{l_1},\tau}(\bs{G}_1)~u_{\bar{\bs{k}}_2+\bs{g}_{l_2},\tau}(\bs{G}_2)~~e^{i\left(\bs{G}_2-\bs{G}_1+\bar{\bs{k}}_2-\bar{\bs{k}}_1+\bs{g}_{l_2}-\bs{g}_{l_1}\right).\bs{r}} \langle c^\dagger_{\bar{\bs{k}}_1+\bs{g}_{l_1},\tau,s} c_{\bar{\bs{k}}_2+\bs{g}_{l_2},\tau,s^\prime}\rangle \nonumber \\
	&=\sum u^\dagger_{\bar{\bs{k}}_1+\bs{g}_{l_1},\tau}(\bs{G}_1)~u_{\bar{\bs{k}}_1+\bs{g}_{l_1}+\bs{g}_{i},\tau}(\bs{G}_2) ~e^{i\left(\bs{G}_2-\bs{G}_1+\bs{g}_{i}\right).\bs{r}} M^{\tau,s,s^\prime}_{\bs{g}_{l},\bs{g}_{i}}(\bar{\bs{k}}_1).\label{eq:rho}
\end{align}
Components of normalized spin vectors, $\bs{S}_{\tau}$, are now can be computed using $S^i_{\tau} = \frac12 \rm{Tr}[\sigma^i \rho_{\tau}]/\rm{Tr}[\rho_{\tau}]$, with $\sigma^i$ the Pauli matrices ($\hbar=1$). In similar manner, the charge density can be written as
$$n_{\tau}(\bs{r}) = \rm{Tr}[\rho_{\tau}(\bs{r})] \equiv \rho_0(\bs{r}) + \delta_{\tau}(\bs{r}),$$
where $\delta_{\tau}(\bs{r})$ is the real-space excess charge relative to the half-filled state and $\rho_0(\bs{r})$ is the charge density at the half-filled state. 

\begin{figure}
	\centering
	\includegraphics[width=0.35\textwidth]{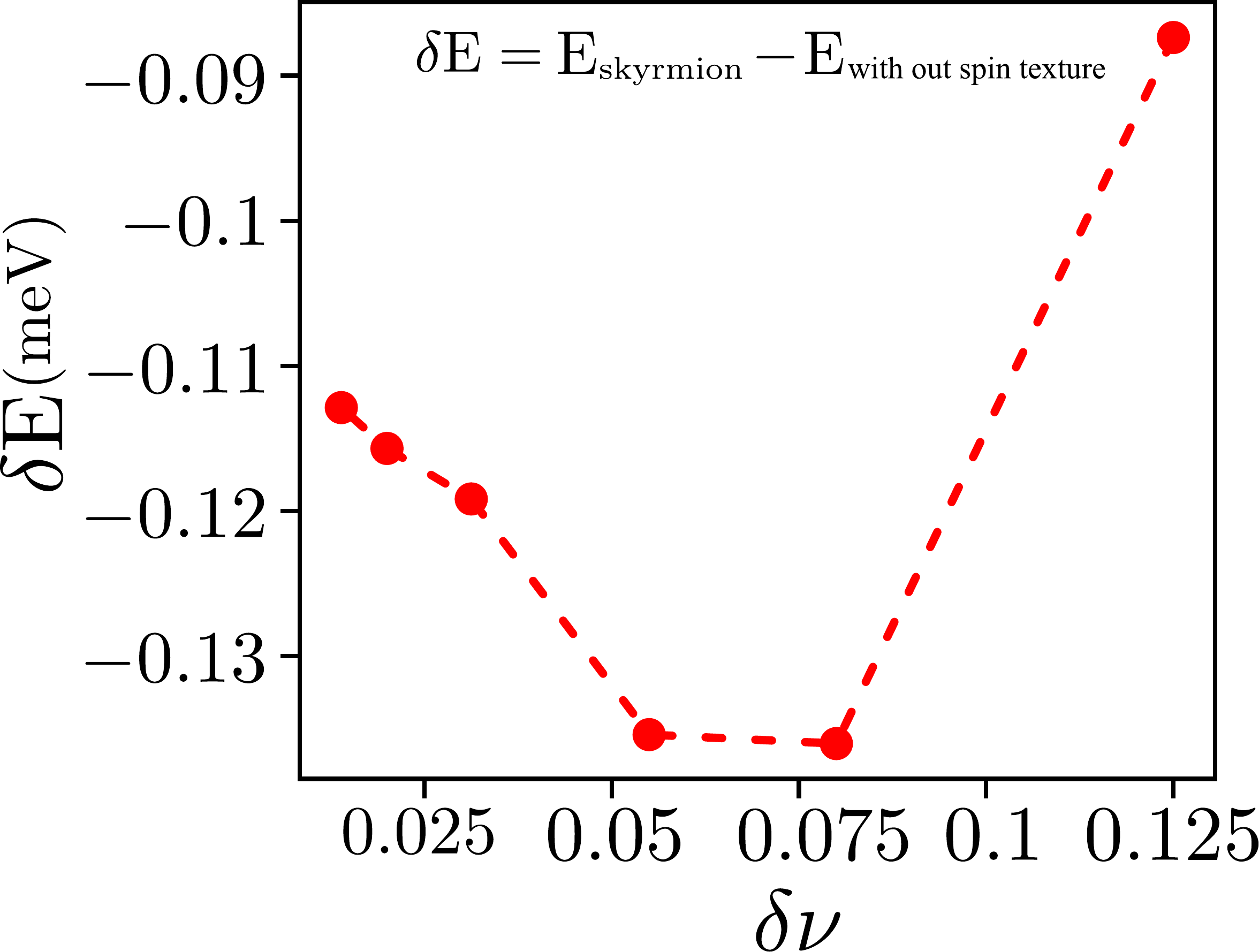}
	\caption{Energy gain ($\delta \rm{E}$) to form a skyrmion stripe state (similar to Fig. 1 of main text) relative to a maximally spin aligned state, as a function of excess filling $\delta \nu$ above half-filling.  Supercell contains 36 \moire unit cells.} \label{fig:energy}
\end{figure}

\subsection{Initial Seeds and Energy Comparison}
In order to search for a skyrmionic ground-state, in the initial iteration we add a non-colinear Zeeman field term to the mean-field Hamiltonian which produces a spin texture in real space, which is spread out over $N_1\times N_2$ unit cells of the original \moire lattice. The magnitude of this term is gradually dropped to zero over subsequent iterations. To produce the triangular skyrme lattice, with a single skyrmion per superlattice unit cell, we add the non-colinear Zeeman term defined by Eq. (\ref{tex1}) to the mean-field Hamiltonian. This is subsequently projected into the Chern band with Chern number $C$. For states with two skyrmions in each supercell, we add the Hamiltonian (projected into the relevant Chern band)
\begin{align}
	h^z_2(\textbf{r}-\textbf{r}_0)&=h^z_1(\textbf{r})-\frac{3}{2}(\sin{\bs{g}_2.(\textbf{r}-\textbf{r}_0)}-\sin{\bs{g}_1.(\textbf{r}-\textbf{r}_0)})\sigma_x +\frac{\sqrt{3}}{2}(\sin{\bs{g}_1.(\textbf{r}-\textbf{r}_0)}+\sin{\bs{g}_2.(\textbf{r}-\textbf{r}_0)}-2\sin{\bs{g}_3.(\textbf{r}-\textbf{r}_0)})\sigma_y\nonumber \\
	&+2(\cos{\bs{g}_1.(\textbf{r}-\textbf{r}_0)}+\cos{\bs{g}_2.(\textbf{r}-\textbf{r}_0)}+\cos{\bs{g}_3.(\textbf{r}-\textbf{r}_0)})\sigma_z\label{tex2}.
\end{align}

To quantify the gain in energy due to forming a spin texture, one may search for a ground-state without one. This is accomplished by starting with an initial seed where the state is maximally spin-polarized even when doped away from half-filling, and no additional non-colinear Zeeman term is added. For the parameter regimes we explored, we find that the resulting spin-aligned states are energetically more costly than the textured states.  In Fig.~\ref{fig:energy} we plot the energy of a skyrme stripe state relative to a spin-aligned state, $\delta {\rm E}$.  The results show that forming the texture is always energetically favorable for these densities.

\subsection{Skyrmion Stripes Without Intervalley Exchange}

\begin{figure*}[t]
	\centering
	\includegraphics[width=0.9\textwidth]{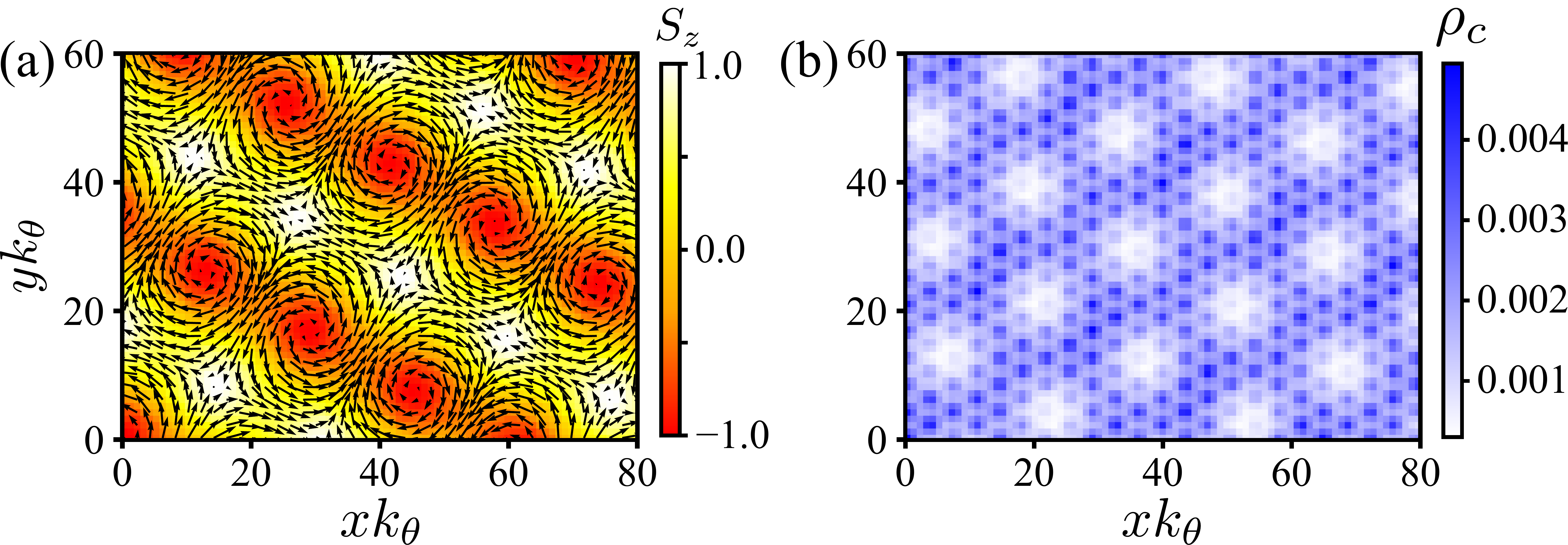}
	\caption{Skyrmion stripe state without intervalley exchange term.
		The supercell of the system is $9\times 8$ times larger than the \moire unit cell and contains two skyrmions per unit cell.  Spin texture is fully contained in one valley while the other is integrally filled.  The Chern number of the non-interacting band is $C=1$. (a): The in plane component of the spin densities (i.e. $S_x$ and $S_y$) are plotted within the 2D plane indicated by the black arrows.  The color bar corresponds to the $S_z$ component of the spin with $S_z=-1$ at the skyrmion center. (b): Excess charge density ($\delta_+ (\bs{r})$) profile of the state relative to half-filled state.  The form of the charge density indicates that state does not truly break $C_3$ symmetry. 
	}\label{fig:without_intervalley_9_8}
\end{figure*}

\begin{figure*}[b]
	\centering
	\includegraphics[width=1.0\textwidth]{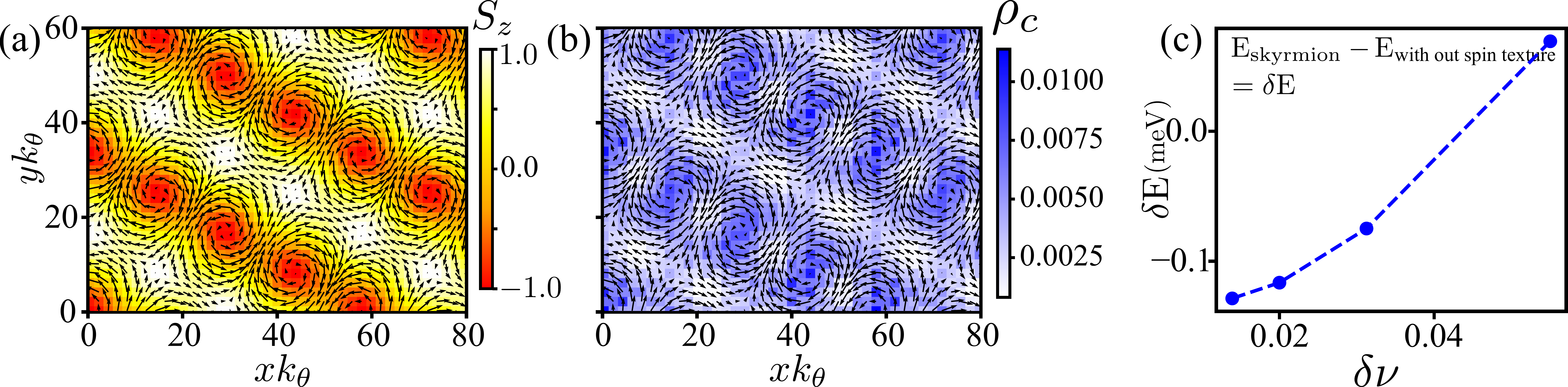}
	\caption{Skyrmion stripe state in a Chern band with Chern number $C=2$. The supercell of the system contains 64 \moire unit cells.  (a): The in plane component of the spin densities ($S_x$ and $S_y$) are plotted within the 2D plane indicated by the black arrows.  The color bar corresponds to the $S_z$ component of the spin with $S_z=-1$ at the skyrmion center. (b): Excess charge density ($\delta_+ (\bs{r})$) profile of skyrme crystal relative to the half-filled state. (c): Energy gain of stripe skyrmion state compared to a maximally spin-polarized state. Here we plot the energy gain ($\delta \rm{E}$) as a function of excess filling, $\delta \nu$ with respect to the half-filling. 
	}\label{fig:skyrmec2}
\end{figure*}

If one excludes the inter-valley coupling term, i.e., the term $H_{\rm{int}}^{(2)}$ in Eq. (\ref{eq:H_t}), the skyrmion stripe state changes in a non-trivial way. An example of the resulting spin and charge density profiles is shown in Fig. \ref{fig:without_intervalley_9_8}. While the ordering in spin appears qualitatively similar to the case with intervalley exchange, no stripe order is apparent in the charge density.  Indeed, this state {\it preserves} $C_3$ symmetry in the following sense: when the spin components of the state are cycled ($S_x \rightarrow S_y$, $S_y \rightarrow S_z$, $S_z \rightarrow S_x$), {\it and} real space vectors are rotated by 120${^\circ}$, the densities in Fig. \ref{fig:without_intervalley_9_8} are unchanged; equivalently, if one cycles the spins but leaves real space vectors unchanged, the stripes apparent in Fig. \ref{fig:without_intervalley_9_8}(a) rotate by 120${^\circ}$.  Thus the stripes in Fig. \ref{fig:without_intervalley_9_8}(a) are apparent only because of the different way in which $S_z$ is represented relative to $S_x$ and $S_y$.  The fact that the charge density, which is a scalar quantity connected to the Pontryagin density, shows no stripe order corroborates the absence of true broken orientational order in this state.

The above observations suggest that breaking the SU(2) symmetry of the spin degree of freedom in the system would lead to true stripe order, which can be observed in the charge density.  We have verified numerically that this is indeed the case, by adding a uniform Zeeman coupling to the Hamiltonian.  In the results presented in the main text, the spontaneously broken spin symmetry (i.e., ferromagnetism) of the integrally filled valley plays precisely the role of such a Zeeman field, through the intervalley exchange term.  In this way the state manifests a very unusual type of spin-orbit coupling arising spontaneously through electron-electron interactions, rather than through any explicit term in the Hamiltonian.


\section{Comparison between partialally valley polarized (PVP) and valley balanced (VB) skyrmion}

The comparison between the ground state energy of  partialally valley polarized (PVP) skyrmion and valley balanced (VB) skyrmion is shown in Table \ref{tab:vp_vu}. We perform the self-consistent Hartree-Fock calculation with a supercell containing 36 \moire unit cell for these two cases. The mean-field Hamiltonian for the two cases is defined by Eq. (\ref{eq:mf_vpp}), (\ref{eq:mf_vpm}) and Eq. (\ref{eq:mf_vc}) respectively. Our findings indicate that skyrme crystal states with an equal distribution of electrons across both valleys possess higher energy than tially parvalley-polarized skyrme crystals.
\begin{center}
	\begin{table}[htbp]
		\centering
		\def\arraystretch{3.0}
		\begin{tabular}{ | c | c | c | c | c | }
			\hline
			$\rm{texture}$& $\rm{order~parameter}$ & $\rm{size~of~skyrme~crystal}$ & $\rm{filling}$ &$\rm{total~energy~of~ground~state}$  \\ \hline
			$\rm{PVP~skyrmion}$ & Eq. (\ref{eq:vpk}) \rm{and} Eq. (\ref{eq:vpkp})   & $6 \times 6$ $\rm{times~larger~of}~$\moire$~\rm{lattice}$ & $0.5+\frac{2}{6\times 6}$ &-12.24 \rm{eV}  \\ \hline
			$\rm{VB~skyrmion}$ & Eq. (\ref{eq:op_vc})  & $6 \times 6$ $\rm{times~larger~of}~$\moire$~\rm{lattice}$ & $0.5+\frac{2}{6\times 6}$ & -11.74 \rm{eV}  \\ \hline
		\end{tabular}
		\caption{Comparison between the ground state energy of PVP skyrmion and VB skyrmion. }\label{tab:vp_vu}
	\end{table}
\end{center}


\section{Skyrmion Stripes in a Chern Number $\bs{C=2}$ Band}
For comparison we computed the Hartree-Fock ground state for a system in which the skyrmion-hosting band has Chern number $C=2$.  As shown in Fig. \ref{fig:band}, such a band can be produced with an appropriate choice of perpendicular electric field.  Typical results are illustrated in Fig.~\ref{fig:skyrmec2}, which are qualitatively quite similar to those of the skyrmion stripe state for a Chern band with $C=1$.  Note in this case each skyrmion hosts a net charge of 2 relative to the integrally filled bands case.

\end{widetext}


\begin{thebibliography}{99}
	
\bibitem{Cao1}
Y. Cao, V. Fatemi, S. Fang, K. Watanabe, T. Taniguchi,
E. Kaxiras, and P. Jarillo-Herrero, Nature, \textbf{556}, 43 (2018).

\bibitem{Cao2}
Y. Cao, V. Fatemi, A. Demir, S. Fang, S.L. Tomarken,
J.Y. Luo, J.D. Sanchez-Yamagishi, K. Watanabe, T.
Taniguchi, E. Kaxiras, et al., Nature, \textbf{556}, 80 (2018).

\bibitem{tbg1}
M. Yankowitz, S. Chen, H. Polshyn, Y. Zhang, K. Watanabe, T. Taniguchi, D. Graf, A.F. Young, and C.R. Dean, Science,\textbf{363}, 1059 (2019).

\bibitem{tbg2}
X. Lu, P. Stepanov, W. Yang, M. Xie, M.A. Aamir, I.
Das, C. Urgell, K. Watanabe, T. Taniguchi, G. Zhang, A.
Bachtold, A.H. MacDonald, and D.K. Efetov, Nature, \textbf{574}, 653 (2019).

\bibitem{koshino1}
Mikito Koshino, \prb~\textbf{99}, 235406 (2019).

\bibitem{koshino2}
J. A. Crosse, Naoto Nakatsuji, Mikito Koshino, and Pilkyung Moon,
Phys. Rev. B \textbf{102}, 035421  (2020).

\bibitem{chebrolu}
Narasimha Raju Chebrolu, Bheema Lingam Chittari, and Jeil Jung, \prb~\textbf{99}, 235417 (2019).


\bibitem{balents}
Balents, L., Dean, C.R., Efetov, D.K. et al. Nat. Phys.  \textbf{16}, 725–733 (2020).

\bibitem{Bistritzer_2011} R. Bistritzer and A.H. MacDonald, PNAS {\bf 108}, 12233-12237 (2011).

\bibitem{tbgcor1}
H. Polshyn, M. Yankowitz, S. Chen, Y. Zhang, K. Watanabe,
T. Taniguchi, C. R. Dean, and A. F. Young, Nat. Phys.  \textbf{15}, 1011 (2019).

\bibitem{tbgcor2}
X. Liu, Z. Wang, K. Watanabe, T. Taniguchi, O. Vafek, and J. Li, Science \textbf{371}, 1261 (2021).

\bibitem{tbgcor3}
Y. Xie, B. Lian, B. Jäck, X. Liu, C.-L. Chiu, K. Watanabe, T. Taniguchi, B. A. Bernevig, and A. Yazdani, Nature \textbf{572}, 101 (2019).

\bibitem{tbgcor4}
Y. Choi, J. Kemmer, Y. Peng, A. Thomson, H. Arora, R. Polski, Y. Zhang, H. Ren, J. Alicea, G. Refael, et al., Nature physics \textbf{15}, 1174 (2019).

\bibitem{tbgcor5} K. P. Nuckolls, M. Oh, D. Wong, B. Lian, K. Watanabe,
T. Taniguchi, B. A. Bernevig, and A. Yazdani, Nature \textbf{588}, 610 (2020).

\bibitem{tbgcor6} Y. Saito, J. Ge, L. Rademaker, K. Watanabe, T. Taniguchi, D. A.
Abanin, and A. F. Young, Nat. Phys. \textbf{17}, 478 (2021).

\bibitem{tbgcor7}
I. Das, X. Lu, J. Herzog-Arbeitman, Z.-D. Song, K. Watanabe, T. Taniguchi, B. A. Bernevig, and D. K. Efetov, Nat. Phys., \textbf{17}, 710 (2021).

\bibitem{tbgcor8}
S. Wu, Z. Zhang, K. Watanabe, T. Taniguchi, and E. Y. Andrei, Nature materials \textbf{20}, 488 (2021).

\bibitem{tbgcor10} P. Potasz, M. Xie, and A. H. MacDonald, Phys. Rev. Lett. \textbf{127},
147203 (2021).

\bibitem{tbgcora}
Jian Kang and Oskar Vafek, Phys. Rev. B \textbf{102}, 035161 (2020).

\bibitem{tbgcorb}
Kasra Hejazi, Xiao Chen, and Leon Balents, Phys. Rev. Research \textbf{3}, 013242 (2021).

\bibitem{tbgcor11}
Xu Zhang, Gaopei Pan, Bin-Bin Chen, Heqiu Li, Kai Sun, and Zi Yang Meng
Phys. Rev. B \textbf{107}, L241105 (2023).

\bibitem{tbgcor12}
Gaopei Pan, Xu Zhang, Hongyu Lu, Heqiu Li, Bin-Bin Chen, Kai Sun, and Zi Yang Meng
Phys. Rev. Lett. \textbf{130}, 016401 (2023).


\bibitem{Po_2019} Hoi Chun Po, Liujun Zou, T. Senthil, and Ashvin Vishwanath, \prb~{\bf 99}, 195455 (2019).


\bibitem{vafek18}
Jian Kang and Oskar Vafek, Phys. Rev. X \textbf{8}, 031088 (2018).

\bibitem{tdbgcr1}
G. William Burg, Jihang Zhu, Takashi Taniguchi, Kenji Watanabe, Allan H. MacDonald, and Emanuel Tutuc, \prl \textbf{123}, 197702 (2019).

\bibitem{tdbgcr2}
Jianpeng Liu, Zhen Ma, Jinhua Gao, and Xi Dai, Phys. Rev. X \textbf{9}, 031021 (2019).

\bibitem{tdbgcr3}
Wang, Y., Herzog-Arbeitman, J., Burg, G.W. et al., Nat. Phys.  \textbf{18}, 48–53 (2022).

\bibitem{tdbgcr4}
Liu, X., Chiu, CL., Lee, J.Y. et al.,  Nat Commun \textbf{12}, 2732 (2021).

\bibitem{tdbgcr5}
Kuiri, M., Coleman, C., Gao, Z. et al. Nat Commun \textbf{13}, 6468 (2022).


\bibitem{lado}
P. Rickhaus, G. Zheng, J. L. Lado, Y. Lee, A. Kurzmann, M. Eich, R. Pisoni, C. Tong, R. Garreis, C. Gold, M. Masseroni, T. Taniguchi, K. Wantanabe, T. Ihn, and K. Ensslin,  Nano Lett. \textbf{19}, 12, 8821–8828 (2019).



\bibitem{Adak_2022} Pratap Chandra Adak, Subhajit Sinha, Debasmita Giri, Dibya Kanti Mukherjee, L. D. Varma Sangani, Kenji Watanabe, Takashi Taniguchi, H.A. Fertig, Arijit Kundu, and Mandar M. Deshmukh, Nature Communications {\bf 13}, 7781 (2022).

\bibitem{mandar2} Sinha, S., Adak, P.C., Surya Kanthi, R.S. et al. Nat Commun \textbf{11}, 5548 (2020).

\bibitem{mandar3}
Sinha, S., Adak, P.C., Chakraborty, A. et al., Nat. Phys.  \textbf{18}, 765–770 (2022).





\bibitem{dassarma}
Fengcheng Wu and S. Das Sarma, \prb~\textbf{101}, 155149 (2020).

\bibitem{tdbgferro1}
C. Shen, N. Li, S. Wang, Y. Zhao, J. Tang, J. Liu,
J. Tian, Y. Chu, K. Watanabe, T. Taniguchi, et al.,  Nat. Phys.  \textbf{16}, 520–525 (2020).

\bibitem{tdbgferro2}
Y. Cao, D. Rodan-Legrain, O. Rubies-Bigordà, J. M.
Park, K. Watanabe, T. Taniguchi, and P. Jarillo-
Herrero,  Nature \textbf{583}, 215–220 (2020).

\bibitem{tdbgferro3}
X. Liu, Z. Hao, E. Khalaf, J. Y. Lee, K. Watan-
abe, T. Taniguchi, A. Vishwanath, and P. Kim,  Nature \textbf{583}, 221–225 (2020).

\bibitem{tdbgferro4}
Le Liu, Xin Lu, Yanbang Chu, Guang Yang, Yalong Yuan, Fanfan Wu, Yiru Ji, Jinpeng Tian, Kenji Watanabe, Takashi Taniguchi, Luojun Du, Dongxia Shi, Jianpeng Liu, Jie Shen, Li Lu, Wei Yang, and Guangyu Zhang,   Phys. Rev. X \textbf{13}, 031015 (2023).

\bibitem{He_2020} He, M., Li, Y., Cai, J. et al.,  Nat. Phys. {\bf 17}, 26-30 (2021).


\bibitem{Zhang_2021} Yang Zhang, Tongtong Liu, and Liang Fu, \prb~{\bf 103}, 155142 (2021).

\bibitem{Padhi_2018} B. Padhi, C. Setty, and Philip W. Phillips, Nano Lett. {\bf 18}, 6175–6180 (2018).

\bibitem{Padhi_2021} B. Padhi, R. Chitra, and Philip W. Phillips,  \prb~{\bf 103}, 125146 (2021).

\bibitem{Li_2021} Li, T., Zhu, J., Tang, Y. et al.,  Nat. Nanotechnol. {\bf 16}, 1068–1072 (2021). 


\bibitem{Skyrmion_book} G.E. Brown and M. Rho, {\it The Multifaceted Skyrmion}, (World Scientific, Singapore, 2010).


\bibitem{Perspectives_book} S. Das Sarma and A. Pinczuk, eds. {\it Perspectives in Quantum Hall Effects,} (John Wiley \& Sons, New York, 1997).

\bibitem{Brey_1995} L. Brey, H. A. Fertig, R. C\^{o}t\'{e}, and A. H. MacDonald,  Phys. Rev. Lett. {\bf 75}, 2562 (1995).

\bibitem{Goerbig21}
Jonathan Atteia, Yunlong Lian, and Mark Oliver Goerbig, \prb~\textbf{103}, 035403 (2021).






\bibitem{Kwan_2022} Yves H. Kwan, Glenn Wagner, Nick Bultinck, Steven H. Simon, and S.A. Parameswaran,  Phys. Rev. X {\bf 12}, 031020.

\bibitem{Bomerich_2020} Thomas B\"omerich, Lukas Heinen, and Achim Rosch, \prb~{\bf 102}, 100408.

\bibitem{Khalaf_2021} E. Khalaf, S. Chatterjee, N. Bultinck, M. P. Zaletel, and A. Vishwanath, Sci. Adv. {\bf 7}, eabf5299 (2021).

\bibitem{Yang_2006} Kun Yang, S. Das Sarma, and A. H. MacDonald,  \prb~{\bf 74}, 075423 (2006).

\bibitem{Doucot_2008} B. Doucot, M. O. Goerbig, P. Lederer, and R. Moessner, \prb~{\bf 78}, 195327 (2008).

\bibitem{Cote_2007} R. C\^ot\'e, D. B. Boisvert, J. Bourassa, M. Boissonneault, and H. A. Fertig, \prb~{\bf 76}, 125320 (2007).

\bibitem{Cote_2008} R. C\^ot\'e, J.-F. Jobidon, and H. A. Fertig, \prb~{\bf 78}, 085309 (2008).

\bibitem{Lian_2016} Y. Lian, A. Rosch, and M.O. Goerbig, Phys. Rev. Lett. {\bf 117}, 056806 (2016).

\bibitem{Bonsall_1977} L. Bonsall and A.A. Maradudin,  \prb~{\bf 15}, 1959 (1977).

\bibitem{nematic}
Rubio-Verdú, C., Turkel, S., Song, Y. et al.,  Nat. Phys.  \textbf{18}, 196–202 (2022).

\bibitem{Fertig_1999} H.A. Fertig, Phys. Rev. Lett. {\bf 82}, 3693 (1999).

\bibitem{Mukhopadhyay_2001} Ranjan Mukhopadhyay, C. L. Kane, and T. C. Lubensky,  \prb~{\bf 64}, 045120 (2001).

\bibitem{Cote_1997} R. C\^ot\'e, A. H. MacDonald, Luis Brey, H. A. Fertig, S. M. Girvin, and H. T. C. Stoof, Phys. Rev. Lett. {\bf 78}, 4825 (1997).

\bibitem{Chaikin_1995} P.M. Chaikin and T.C. Lubensky, {\it Principles of Condensed Matter Physics}, (Cambridge University Press, New York, 1995).

\bibitem{Auerbach_1994} Assa Auerbach, {\it Interacting Electrons and Quantum Magnetism}, (Springer-Verlag, New York, 1994).

\bibitem{Fu_2021} Hailong Fu , Ke Huang, Kenji Watanabe , Takashi Taniguchi, and Jun Zhu, Phys. Rev. X {\bf 11}, 021012 (2021).

\bibitem{Wei_2018} D. S. Wei, T. van der Sar, S. H. Lee, K. Watanabe,
T. Taniguchi, B. I. Halperin, and A. Yacoby, Science {\bf 362}, 229 (2018).

\end{thebibliography}
\end{document}